\documentclass{aa}  

\usepackage{txfonts}
\usepackage[export]{adjustbox}
\usepackage{subcaption}
\usepackage[colorlinks,linkcolor=blue,anchorcolor=blue,citecolor=blue]{hyperref}
\newcommand{\sfrd}{$\Sigma_{\rm SFR}$}

\newcommand{\usfrd}{$\rm M_{\odot}\,yr^{-1}\,kpc^{-2}$}
\newcommand{\udif}{$\rm cm^2\,s^{-1}$}

\begin{document} 

    \title{Nearby galaxies in the LOFAR Two-metre Sky Survey}

   \subtitle{III. Influence of cosmic-ray transport on the radio--SFR relation}

   \author{V. Heesen\inst{1}
          \and
          S.~Schulz\inst{1}
          \and
          M.~Br\"uggen\inst{1}
          \and
          H.~Edler\inst{1}
          \and
          M.~Stein\inst{2}
          \and 
          R.~Paladino\inst{3}
          \and
          A.~Boselli\inst{4}
          \and
          A.~Ignesti\inst{5}
          \and
          M.~Fossati\inst{6,7}
          \and
          R.-J.~Dettmar\inst{2}
         }

   \institute{Hamburg University, Hamburger Sternwarte, Gojenbergsweg 112, 21029 Hamburg, Germany\\
              \email{volker.heesen@hs.uni-hamburg.de}
        \and
        Ruhr University Bochum, Faculty of Physics and Astronomy, Astronomical Institute (AIRUB), 44780 Bochum, Germany
        \and
        INAF, Istituto di Radioastronomia, Via Gobetti 101, I-40129 Bologna, Italy
        \and 
        Aix-Marseille Univ., CNRS, CNES, LAM, Marseille, France
        \and
        INAF-Padova Astronomical Observatory, Vicolo dell’Osservatorio 5, I-35122 Padova, Italy
        \and
        Universit\'a di Milano-Bicocca, piazza della scienza 3, 20100 Milano, Italy
        \and 
        INAF - Osservatorio Astronomico di Brera, via Brera 28, 20121 Milano, Italy}

   \date{Received date / Accepted date}
   
 
  \abstract
   {In order to understand galaxy evolution, it is essential to measure star formation rates (SFRs) across Cosmic times.}
   {The use of radio continuum emission as an extinction-free star formation tracer necessitates a good understanding of the influence of cosmic-ray electron (CRE) transport that we are aiming to improve with this work.}
   {We analyse the spatially resolved radio continuum--star-formation rate (radio--SFR) relation in 15 nearby galaxies using data from the LOw Frequency ARray (LOFAR) and the Westerbork Synthesis Radio Telescope (WSRT) at 144 and 1365\,MHz, respectively. The hybrid SFR maps are based on observations with {\it Spitzer} at 24\,$\muup$m and with {\it GALEX} at 156\,nm. Our pixel–by–pixel analysis at $1.2$\,kpc resolution reveals the usual sublinear radio--SFR relation for local measurements which can be linearised with a smoothing experiment, convolving the hybrid SFR map with a Gaussian kernel that provides us with the CRE transport length.}
   {CRE transport can be described as energy-independent isotropic diffusion. If we consider only young CREs as identified with the radio spectral index, we find a linear relation showing the influence of cosmic-ray transport. We then define the CRE calorimetric efficiency as the ratio of radio-to-hybrid SFR surface density and show that it is a function of the radio spectral index. If we correct the radio--SFR relation for the CRE calorimetric efficiency parametrised with the radio spectral index, it becomes nearly linear with a slope of $1.01\pm 0.02$ independent of frequency.}
   {The corrected radio--SFR relation is universal and holds, both, for global and local measurements.}



\titlerunning{Radio continuum and star formation in nearby galaxies}
\authorrunning{V.~Heesen et al.}

   \maketitle
%

\section{Introduction}

Cosmic rays and magnetic fields are important ingredients for understanding galaxy evolution. Both have energy densities comparable to the thermal gas and so influence the dynamics of gas accretion and outflows. Cosmic rays are now suspected to play an important role in galactic winds and the dynamics of the circumgalactic medium \citep[e.g.][]{pakmor_20a,van_de_voort_21a}. Magnetic fields regulate the transport of cosmic rays and thus they need to be taken into account as well; but they are crucial of their own such as playing an important role during the early stages of star formation via regulating the transport of angular momentum during the collapse of molecular clouds. The number density of cosmic rays can be approximated by a power law reaching from GeV above many orders of magnitude, although more recent works have shown deviations from it in particular at energies in excess of PeV \citep[see][for a review]{gaisser_13a}. While cosmic rays may reach energies beyond even EeV, GeV-cosmic rays are dynamically most important for gas motions as they contain the bulk of the cosmic-ray energy density. It is these cosmic rays that we can observe via their electron component the cosmic-ray electrons (CREs). The same electrons emit synchrotron emission while spiralling around magnetic field lines allowing us to observe them in the radio continuum.

Radio continuum emission is the result of massive star formation. Stars of $O$ and $B$ spectral classification ionise hydrogen in their environment forming Str\"omgren spheres. This ionised gas is visible in the radio continuum via free--free emission of electrons deflected in the electric field of protons; this radiation is referred to as thermal emission. At gigahertz and lower frequencies, the second emission process is even more important and dominates at fractions of $>90\,\%$ in late-type galaxies \citep{tabatabaei_17a,stein_23a}. Massive stars end their relatively short lifes in core-collapse supernovae. In the aftermath of that explosion, strong shocks are formed that are able to accelerate cosmic rays in supernova remnants as predicted by theory \citep{drury_83a} and confirmed by observations \citep{aharonian_07a}. In addition to these primary cosmic rays, hadronically interacting cosmic-ray protons with ambient gas generate charged pions that decay into secondary electrons and positrons which also radiate synchrotron emission. These cosmic rays are able to diffuse in galaxies, while emitting synchrotron radiation from their electrons; this radiation is referred to as non-thermal emission. In summary, the radio continuum luminosity of a galaxy should be proportional to the number of massive stars. This would allow one to derive the star-formation rate (SFR) of a galaxy by extrapolating from the massive stars that must have been recently formed giving rise to the radio continuum--star-formation rate (radio--SFR) relation \citep{condon_92a,bell_03a,murphy_11a}.

The radio--SFR relation has been extensively studied in galaxies \citep{boselli_15a,li_16a,davies_17a,tabatabaei_17a,gurkan_18a,hindson_18a,smith_21a,heesen_22a}. For integrated, global studies the slope of the relation appears to be slightly super-linear, so that galaxies with higher SFRs have radio luminosities $L_\nu$ that are higher in comparison with their SFR. This appears to be particularly the case at lower frequencies of hundreds of megahertz with the relation $L_\nu\propto {\rm SFR}^{1.4-1.5}$ observed, where the escape of CRE plays an important role due to the longer electron lifetimes \citep{heesen_22a}. In contrast, the spatially resolved radio--SFR relation is significantly sub-linear. This means that areas with low SFR surface densities are bright in the radio continuum emission, whereas the opposite is the case for areas with high SFR surface densities. This is usually ascribed to cosmic-ray transport as well, where the CREs diffuse away from star-forming areas into the galaxy outskirts and inter-arm regions \citep{berkhuijsen_13a,heesen_14a}. Indeed the radio continuum map of a galaxy can be approximated as the smoothed out version of the SFR map where the size of the smoothing kernel can be associated with the CRE transport length \citep{vollmer_20a,heesen_23b}.

Also it has emerged that the age of the CRE is an important secondary parameter. If one restricts the analysis of the radio--SFR relation to areas with high SFR surface densities, the relation in individual galaxies becomes almost linear \citep{dumas_11a,basu_12a}. Such a linear relation is predicted if the galaxy is an electron calorimeter and the CREs lose all their energy before being transported away. This means that the radio--SFR relation for young CRE found near their sources (star-forming regions) is not yet affected by cosmic-ray transport. \citet{heesen_19a} found tentative evidence that the radio--SFR relation is universal and close to linear across several galaxies if one considers only areas with radio spectral indices indicative of young CRE. However, they considered only four galaxies limiting their applicability of conclusions. Hence, what is thus far missing is a study of the spatially resolved radio--SFR relation in a representative sample of nearby galaxies. This has now become possible with data from the LOFAR Two-metre Sky Survey \citep[LoTSS;][]{shimwell_17a,shimwell_19a,shimwell_22a}. \citet{heesen_22a} defined a sample of 76 nearby galaxies in LoTSS that forms the basis of our analysis.

Our aim is to measure star formation rates in nearby galaxies using radio continuum emission. There are two main motivations behind this: first, the existence of a universal radio--SFR relation, both, between galaxies and within galaxies implies a direct link between radio emissions and star formation. Second, if such a link was established it could be used to measure SFRs in galaxies of the local Universe \citep{beswick_15a}. The fact that radio emission is extinction-free as opposed to ultraviolet, H\,$\alpha$ and partially also mid-infrared based methods would allow astronomers to measure star formation rates in high redshift galaxies. On the other hand, radio continuum-based methods could be impaired by inverse Compton losses that become more important at higher redshifts \citep{murphy_09a}. The use of LOFAR is especially helpful in this context. At lower frequencies the thermal component of the radio emissions is even smaller with fractions below 5\,\% \citep{stein_23a}. Lower frequencies also allow us to inspect lower energy CREs. Since the CREs travel significant distances while synchrotron radiating the resulting radio map is be a smeared out version of the true star formation distribution with the effect even more pronounced at lower frequencies \citep{heesen_19a}.

This paper is organised as follows: in Section~\ref{ss:context} we provide some context how we can measure cosmic-ray transport with radio continuum data. In Section~\ref{s:data_and_methodology}, we present the data we have used and our methodology. Section~\ref{s:results} summarises our results. In Section~\ref{sec:discussion} we present our discussion while we present our conclusions in Section~\ref{sec:conclusions}.

\subsection{Measuring cosmic-ray transport in nearby galaxies}
\label{ss:context}

There are broadly speaking three ways of measuring cosmic-ray transport parameters in nearby galaxies. First, one can compare the luminosity of a galaxy either in the radio continuum or in the gamma rays with the SFR. These correlations are slightly superlinear which can be interpreted that galaxies are better cosmic-ray calorimeters at higher masses and sizes \citep{ackermann_12a,li_16a,heesen_22a}.  Second, one can use spectral analyses. If the spectrum is a continuous power-law in agreement with the injection spectrum, one can deduce that energy-independent transport such as by advection is the dominant transport. This is observed in some galaxies, both, in gamma rays \citep{abramowski_12a} as well as in the radio continuum \citep{kapinska_17a}. However, in the latter case the interpretation is compounded by the effect of various energy losses that tend to flatten the spectrum at low energies via ionisation losses and steepen the spectrum at high energies via synchrotron and inverse Compton losses. Thirdly, one can compare the distribution of cosmic rays with the distribution of their sources. This method has been employed in gamma rays \citep{murphy_12a} and more widely in the radio continuum \citep{murphy_08a,vollmer_20a}. Results so far also seem to favour energy-independent transport via diffusion, streaming or advection \citep[e.g.][]{heesen_23a, stein_23a}.

Most cosmic-ray transport studies have concentrated on the Milky Way, where detailed cosmic-ray spectra can be measured. Here, we review some of the spectral properties of nearby galaxies found in the radio continuum.  The radio continuum spectrum of nearby galaxies indicates a non-thermal radio spectral index of $\approx$$-1.0$ \citep{tabatabaei_17a} which corresponds to an electron spectral index of $\gamma=3.0$.\footnote{Assuming the cosmic-ray number density is $N(E)\propto E^{-\gamma}$ where $E$ is the energy. The (non-thermal) radio spectral index is then $\alpha=(1-\gamma)/2$ where the radio continuum intensity scales with frequency as $I_\nu\propto \nu^\alpha$.} This is a slightly flatter spectrum than what Galactic cosmic-ray electrons with energies 40--300\,GeV show with $\gamma\approx 3.1$ \citep{adriani_18a}, including results from the Calorimetric Electron Telescope (CALET), Alpha Magnetic Spectrometer (AMS-02), {\it Fermi} Large Area Telescope ({\it Fermi}-LAT), and Dark Matter Particle Explorer (DAMPE).  In the Galaxy, the cosmic-ray spectrum flattens below 10\,GeV. This can be attributed to a change from advection-dominated transport for low energies to diffusion-dominated transport at higher energies \citep{blasi_12a}. At higher energies of $>$10\,GeV the spectrum of cosmic-ray nuclei deviates from that of the protons where the proton spectrum tends to be about $\delta \gamma=0.1$--0.2 steeper. Such a difference could be related either to the accelerating sources themselves or to a transport-related break \citep{becker_tjus_20a}.

At  the low energies of a few GeV that we observe in the radio continuum we can assume that all cosmic rays are of Galactic origin and related to star formation \citep{gaisser_13a}. The most likely source class is that of supernova remnants where the radio spectral index is $-0.5$, corresponding to $\gamma=2.0$ \citep{ranasinghe_23a}. Hence, we find significant evidence for spectral ageing as the spectrum is much steeper than the injection spectrum. It is particularly interesting to learn about the transport of cosmic-ray electrons as they may allow us to identify individual accelerators given their short transport distances, in particular at energies in excess of 1\,TeV \citep{mertsch_18a}. On the other hand, the high energy losses of electrons complicates their transport analysis.

\section{Data and methodology}
\label{s:data_and_methodology}

\subsection{Data}
\label{ss:data}
Our galaxy sample consists of 15 galaxies that were included in LoTSS-DR2 and in SINGS \citep[The SIRTF Nearby Galaxies Survey;][]{kennicutt_03a}. These maps at 144~MHz are available at resolutions of $6\arcsec$ and $20\arcsec$. The sample includes also galaxies that were already observed by LoTSS but not yet publically released. These galaxies were observed and analysed in the same way as for LoTSS-DR2. Some of the physical properties can be found in Table~\ref{tab:sample}. The hybrid SFR maps are created by a combination of \emph{GALEX} far-UV 156\,nm and \emph{Spitzer} mid--IR $24\,\muup$m data \citep{leroy_08a}. Higher frequency radio maps are taken from the SINGS survey conducted at the Westerbork Synthesis Radio Telescope \citep[WSRT--SINGS;][]{braun_07a, heald_09a} with an observing frequency of 1365\,MHz. For NGC~4254 we use a $1.4$-GHz map from the MeerKAT telescope instead of WSRT that was obtained by \citet{edler_23a}. Although this galaxy is included in WSRT--SINGS, the MeerKAT map has superiour resolution and sensitivity.

Additional work was done on all of the maps. They were changed to have the same gridding (in pixel per arcsec) and in size (in pixel). Background sources such as active galactic nuclei (AGN) were masked. Finally the maps were convolved to a circular Gaussian beam to change their resolution to $1.2$\,kpc where this was possible with a few exceptions. We  adopt a  spatial resolution of $1.2$\,kpc because it is the maximum achievable resolution possible for all our galaxies where the limiting factor is the angular resolution of the WSRT. In the few cases where this was not possible due to the distance of the source or the angular resolution of the map, the maps were convolved to the nearest possible circular beam size. \citet{heesen_14a} tested resolutions of $0.7$ and $1.2$\,kpc and found little impact on their results.


\begin{table*}
	\centering
	\caption{Properties of galaxies in the sample.}
	\label{tab:sample}
        \begin{tabular}{l ccccc ccc} 
          \hline\hline
          
          Galaxy & $d$ & $r_\star$ & \sfrd & $B_{\rm eq}$  & $p$ & $v$ & $i$ & SFR  \\ 
   & (Mpc) & (kpc) & (\usfrd) & ($\muup\rm G$) & (\%) & $(\rm km\,s^{-1})$ & $(\degr)$ & $(\rm M_\sun\,yr^{-1})$  \\
       (1) & (2)  & (3) & (3) & (4) & (5) &  (6) & (7) & (8)\\ \hline
    NGC 0628 & $7.2$  & $11.5$ & $2.8\times 10^{-3}$ & $9.0$    & $11.50$ & 217 &  7 & 1.19 \\ 
    NGC 0925 & $9.1$  & $14.9$ & $1.1\times 10^{-3}$ & $6.2$    & $0.67$ & 136 & 66 & 0.75 \\ 
    NGC 2403 & $3.06$ & $8.0$  & $4.5\times 10^{-3}$ & $7.5$    & $7.78$ & 134 & 63 & 0.90 \\ 
    NGC 2841 & $14.1$ & $16.0$ & $2.3\times 10^{-3}$ & $6.4$    & $5.80$ & 302 & 74 & 1.88 \\
    NGC 2903 & $8.9$  & $13.2$ & $1.1\times 10^{-2}$ & $11.0$   & $3.04$ & 190 & 66 & 5.96  \\
    NGC 2976 & $3.55$ & $3.1$  & $3.9\times 10^{-3}$ & $8.3$    & $4.41$ & 92  & 54 & 0.12 \\                   
    NGC 3184 & $11.7$ & $14.0$ & $2.4\times 10^{-3}$ & $7.5$    & $6.25$ & 210 & 16 & 1.49 \\ 
    NGC 3198 & $14.1$ & $14.0$ & $2.4\times 10^{-3}$ & $6.0$    & $1.02$ & 150 & 72 & 1.49 \\
    NGC 3938 & $17.9$ & $15.6$ & $3.9\times 10^{-3}$ & $8.9$    & $7.75$ & 128 & 36 & 2.98 \\
    NGC 4254 & $14.4$ & $10.5$ & $1.7\times 10^{-2}$ & $16.0$   & $4.71$ & 300 & 33 & 5.94 \\
    NGC 4725 & $11.9$ & $13.2$ & $2.2\times 10^{-3}$ & $6.5$    & $3.60$ & 257 & 43 & 1.19 \\
    NGC 4736 & $4.66$ & $4.9$  & $1.2\times 10^{-2}$ & $12.1$   & $3.59$ & 156 & 41 & 0.94 \\
    NGC 5055 & $7.94$ & $15.2$ & $5.9\times 10^{-3}$ & $9.1$    & $3.89$ & 192 & 59 & 2.98 \\ 
    NGC 5194 & $8.0$  & $16.0$ & $1.5\times 10^{-2}$ & $12.0$   & $5.70$ & 219 & 20 & 11.89 \\ 
    NGC 7331 & $14.5$ & $22.8$ & $4.6\times 10^{-3}$ & $8.0$    & $2.45$ & 244 & 77 & 7.48 \\
    \hline
    \end{tabular}
    \tablefoot{Column (2) $d$ is the distance \citep{heesen_22a}; (3) $r_\star$ is the star formation radius as obtained from radio continuum  \citep{heesen_22a}; (4) $B_{\rm eq}$ is the total magnetic field strength estimated from energy equipartition  \citep[][for NGC\,2903 we use the $B$--\sfrd relation]{chyzy_07a,mulcahy_17a,heesen_23a}; (5) $p$  is the polarisation fraction at 1365 MHz  \citep{heald_09a}; (6) $v$ is the rotation velocity of the galaxy  \citep{heesen_22a}; (7) $i$ is the inclination angle  \citep{heesen_22a}; and (8) SFR is the integrated star formation rate  \citep{heesen_22a}.
    }
\end{table*}

\subsection{Radio star formation rates}
\label{ss:radio_sfr}

The radio--SFR relation connects the radio continuum luminosity with the SFR for integrated i.e. global measurements. In the spatially resolved case we deal with  SFR surface density $\Sigma_{\text{SFR}}$ and radio intensity $I_{\nu}$ instead. To this end we adapt the widely used Condon-relation \citep{condon_92a}, although similar more recent incarnations exist as well \citep{murphy_11a}. Condon’s relation assumes a Salpeter initial mass function (IMF) to extrapolate from the number of massive stars ($M>5\,{\rm M_\sun}$) that show up in the radio continuum to that of all stars formed ($0.1 < M/{\rm M_\sun} < 100$). The hybrid SFR maps are based on a broken power-law IMF \citep{calzetti_07a}, so their derived SFRs are a factor of $1.59$ lower than using the Salpeter IMF. We have thus scaled Condon’s relation in this paper accordingly:
\begin{equation}
    \left. \frac{\rm SFR_{\text{RC}}}{\rm M_\sun\,yr^{-1}} \right|_{> 0.1 {\rm M_\sun} } = 0.75 \times 10^{-21} \frac{L_{1.4\,\rm GHz}}{\rm W\,Hz^{-1}} .
    \label{eq:condon}
\end{equation}
Hence, the radio continuum SFR surface density is \citep{heesen_14a,heesen_19a}
\begin{equation}
    \left. \frac{(\Sigma_{\text{SFR}})_{\text{RC}} }{\rm M_\sun\,yr^{-1} \, kpc^{-2}} \right|_{> 0.1 {\rm M_\sun} } = 8.8 \times 10^{-8} \frac{I_{1.4\,\rm GHz}}{\rm Jy\,sr^{-1}} \,.
    \label{eq:condon_resolved}
\end{equation}
In our data the intensity is measured in Jy beam$^{-1}$. To convert from Jy sr$^{-1}$ we use the FWHM (full width at half--maximum) of the synthesized beam measured in arcsec. The solid angle subtended by a Gaussian beam relates to its area as $\Omega = 2.66 \times 10^{-11} \times \rm FWHM^2\,sr$. We will also be using the Condon-relation at different radio frequencies. Thus we scale with frequency $\nu$ relative to $1.4\,\rm GHz$ assuming a simple power law $I_{\nu} \propto \nu ^{\alpha}$ with a radio spectral index of $\alpha=-0.8$. This is of course a simplification as the radio continuum spectrum cannot be described with a single power-law but there is spectral curvature \citep{chyzy_18a}. Also this spectral index applies to galaxies at gigahertz frequencies for the total radio continuum emission \citep[including both the thermal and non-thermal emission;][]{tabatabaei_17a} whereas the spectrum is flatter at lower frequencies \citep{marvil_15a}. We note that the choice of the radio spectral index does not influence our results since the calibration is relative: Equation\,\eqref{eq:condon} is our definition and we then calibrate any deviations from it. 

Our final modified version of the Condon-relation now reads
\begin{equation}
    \frac{(\Sigma_{\text{SFR}})_{\text{RC}} }{\rm M_\sun\,yr^{-1} \,kpc^{-2}} = \left( \frac{\nu}{1.4\,\rm GHz} \right)^{0.8} \, \left( \frac{\text{FWHM}}{\text{arcsec}} \right)^{-2} \frac{3310 I_{\nu}}{\rm Jy\, beam^{-1}} \,.
    \label{eq:condon_final}
\end{equation}
We also calculated radio spectral indices as
\begin{equation}
    \alpha = \frac{\log \left( \frac{I_1}{I_2} \right)}{\log \left( \frac{\nu_1}{\nu_2} \right)}\,.
\end{equation}
with corresponding uncertainties \citep[see][for details]{heesen_22a}. 

Our analysis was done using the software \texttt{radio--pixel--plots} ({\sc rpp})\footnote{\url{github.com/sebastian-schulz/radio-pixel-plots}}.  As input, it requires two radio maps and one hybrid \sfrd\ map, prepared as described above and calculates the spatially resolved (pixel by pixel) SFR surface densities from the radio maps using the Condon relation. Then it maps the results pixel-by-pixel onto the SFR map, performs a double logarithmic linear fit and plots the result. 

In order to calculate the map noise $\sigma_{\rm rms}$ a box-shaped region is defined in an area where no sources are visible. Furthermore, the pixels from the original image are averaged into new, larger ones with a size of 1.2\,kpc. Let $d$ be the distance to the galaxy in Mpc, $h$ the side length of an  original pixel in arcsecond and $r$ the desired resolution (the new pixel size) in pc, then the ratio of original-to-new pixels along one image side
\begin{align}
     \frac{n_{\text{original}}}{n_{\text{new}}} = \frac{r}{ 10^6 d  \tan ( 1\arcsec) h}\,.
\end{align} 
{\sc rpp} does allow this ratio to have non integer values. If an original pixel is only partially inside a new pixel, it contributes to the new value proportionally to the surface area inside the new pixel. After the resolution is converted to $1.2$\,kpc, we apply a $3 \sigma$ cutoff. The resulting error is 
\begin{align}
    \sigma = \sqrt{\sigma_{\rm rms}^2 + \left( \epsilon I_\nu  \right)^2} \,,
\end{align}
where $\epsilon=0.05$ is the calibration uncertainty. This error underestimates the absolute error of the LOFAR data which is about 10\,\% \citep{shimwell_19a}, but this smaller error, dominated by the uncertainties in the deconvolution, is a good estimate, because we are most interested in the spatial variation across maps. Since we need all three values (two radio and one SFR), any pixel that is below the 3\,$\sigma$ threshold will be removed from maps. Finally, the coarse radio map is converted into SFR surface densities using the Condon-relation (Eq.\,\ref{eq:condon_final}) and then mapped onto the SFR map. SFR surface densities from radio continuum and the hybrid data are expected to follow a linear relation in a double logarithmic plot 
\begin{equation}
    \log_{10} \left[ (\Sigma_{\text{SFR}})_{\text{RC}} \right] = a \log_{10} \left[ (\Sigma_{\text{SFR}})_{\text{hyb}} \right] + b \,, \label{eq:loglog_fit}
\end{equation}
where $(\Sigma_{\text{SFR}})_{\text{hyb}}$ denotes the hybrid SFR surface density. This is fitted to our data using the orthogonal distance regression ({\sc odr}) package which is part of the {\sc SciPy} Python library. It implements a \textsc{fortran} algorithm of the same name and uses a modified Levenberg--Marquardt algorithm. It has the advantage of being able to deal with both $x$ and $y$ errors.

\begin{figure*}
    \centering
     \begin{subfigure}[t]{0.03\linewidth}
        \textbf{(a)}    
    \end{subfigure}
    \begin{subfigure}[t]{0.46\linewidth}
    \includegraphics[width=\linewidth,valign=t]{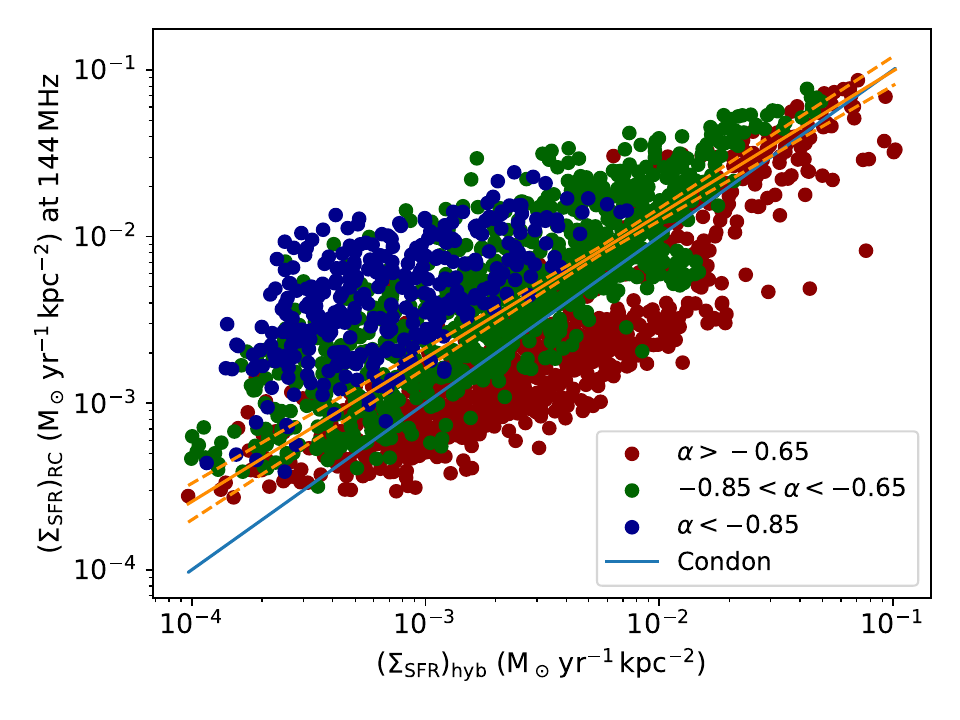}
    \end{subfigure}
    \begin{subfigure}[t]{0.03\linewidth}
        \textbf{(b)}    
    \end{subfigure}
    \begin{subfigure}[t]{0.46\linewidth}
         \includegraphics[width=\linewidth,valign=t]{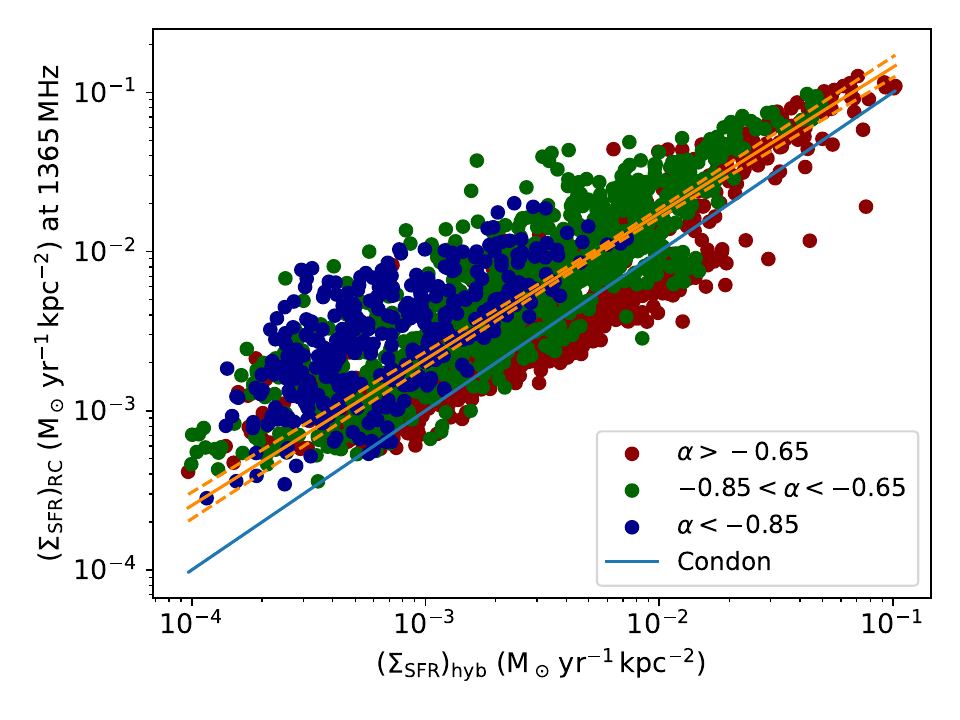}
    \end{subfigure}
    \\
     \begin{subfigure}[t]{0.03\linewidth}
        \textbf{(c)}    
    \end{subfigure}
    \begin{subfigure}[t]{0.46\linewidth}
    \includegraphics[width=\linewidth,valign=t]{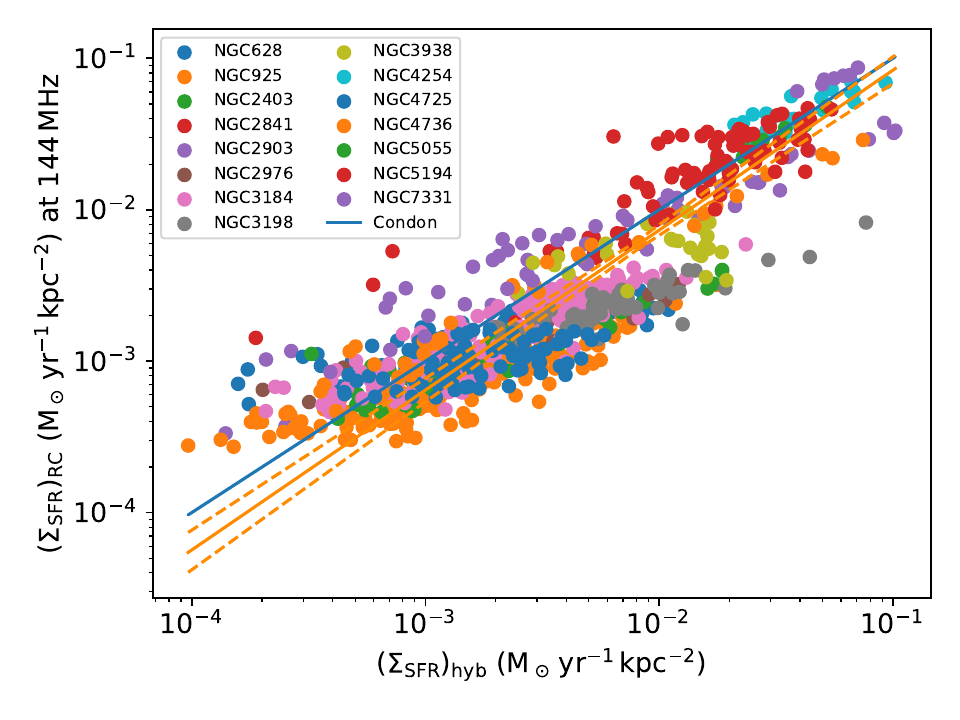}
    \end{subfigure}
    \begin{subfigure}[t]{0.03\linewidth}
        \textbf{(d)}    
    \end{subfigure}
    \begin{subfigure}[t]{0.46\linewidth}
         \includegraphics[width=\linewidth,valign=t]{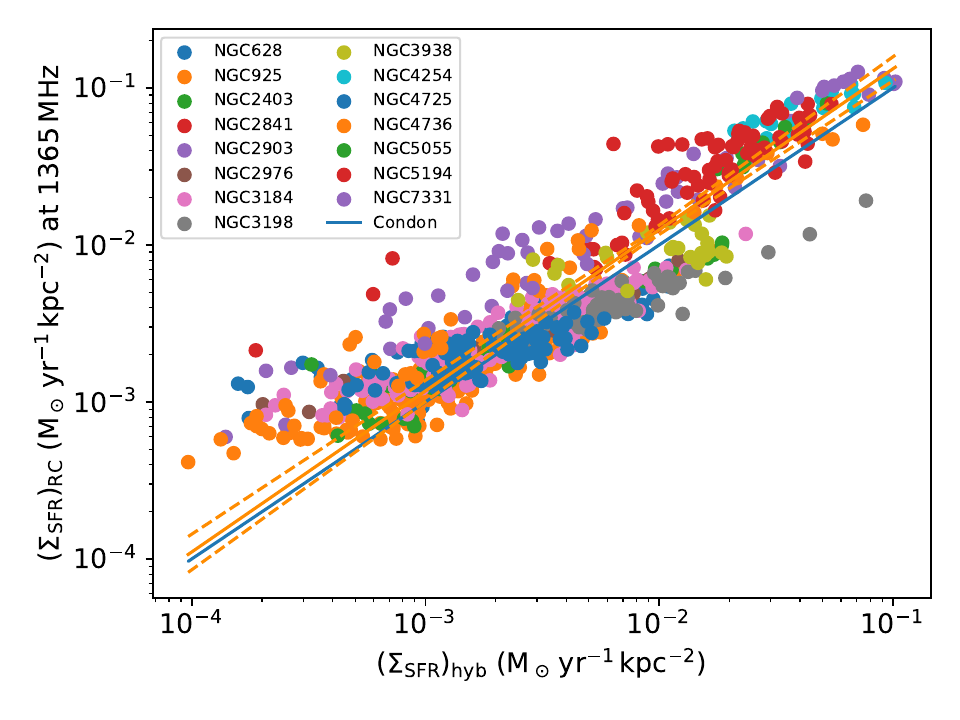}
    \end{subfigure}
    \caption{Combined radio–SFR plots. Panels (a) and (b) show the radio \sfrd\ at 144 and 1365\,MHz, respectively, as function of the hybrid \sfrd. Data points are colour-coded, so that young CREs are shown in red, middle-aged CREs are shown in green, and old CREs are shown in blue. Panels (c) and (d) show the relation only for young CREs, where the various galaxies are now colour coded. Solid lines show in blue the 1:1 Condon relation and in orange the best-fitting power-law relations with 3\,$\sigma$ confidence intervals indicated by dashed lines. Details of the best-fitting relations can found in Tab.\,\ref{tab:diff}.}
    \label{fig:radio_sfr}
\end{figure*}

\subsection{Smoothing experiment}
\label{sec:smoothing}
As in \citet{heesen_23b} we performed a smoothing experiment using {\sc rpp} in order to measure the CRE transport length. To this end we linearise the spatially resolved radio--SFR relation by convolving the hybrid SFR maps with a Gaussian kernel. The choice of kernel is motivated by our assumption that diffusion and not streaming or advection is the mechanism for CRE transport in galactic discs unlike in galactic haloes \citep[e.g.][]{stein_23a}. Otherwise, the kernel would be exponential instead. We define the CRE transport length $l_{\text{CRE}}$ as half of the FWHM of the Gaussian kernel.  Following \citet{heesen_14a}, we define the transport length equal to the diffusion length and the conversion from FWHM is:
\begin{align}
    l_{\text{CRE}} = \frac{1}{2} ( 2 \sqrt{2 \ln 2}) \sigma_{xy} \approx 1.177  \sigma_{xy},
\end{align}
where $\sigma_{xy}$ is the standard deviation of the Gaussian kernel. In order to estimate the uncertainty of $l_{\rm CRE}$ we assumed that the slope of the radio--SFR relation is $1.00\pm 0.05$, where the uncertainty of the slope will then be translated into the uncertainty of $l_{\rm CRE}$ \citep{heesen_19a}. Further details of this method can be found in \citet{heesen_23a}.

Above experiment assumes isotropic diffusion employing a circular kernel. However, we also tested the possible influence of anisotropic diffusion. Several galaxies in our sample have inclination angles of $i > 60\degr$. If we assume anisotroic diffusion in the disc plane following the plane-parallel magnetic field,  the projected diffusion length along the minor axis would have to be corrected with factors of $\cos(i)<1/2$. To test whether this would change our results we modified the convolution kernel to account for the geometric projection effect. The \texttt{Gaussian2DKernel} can be used as an elliptical kernel by providing a standard deviation for each coordinate direction: $\sigma_x$ and $\sigma_y$. In terms of the previous standard deviation $\sigma_{xy}$ we define them as
\begin{align}
    \sigma_x = \sigma_{xy} \,\,\, \text{and}\,\,\,
    \sigma_y = \sigma_{xy} \cos(i).
    \label{eq:inclination}
\end{align}
The orientation of the ellipse can be changed through an additional angle $\theta$ which is the position angle of the major axis.


\begin{figure*}
    \centering
     \begin{subfigure}[t]{0.03\linewidth}
        \textbf{(a)}    
    \end{subfigure}
    \begin{subfigure}[t]{0.46\linewidth}
        \includegraphics[width=\linewidth,valign=t]{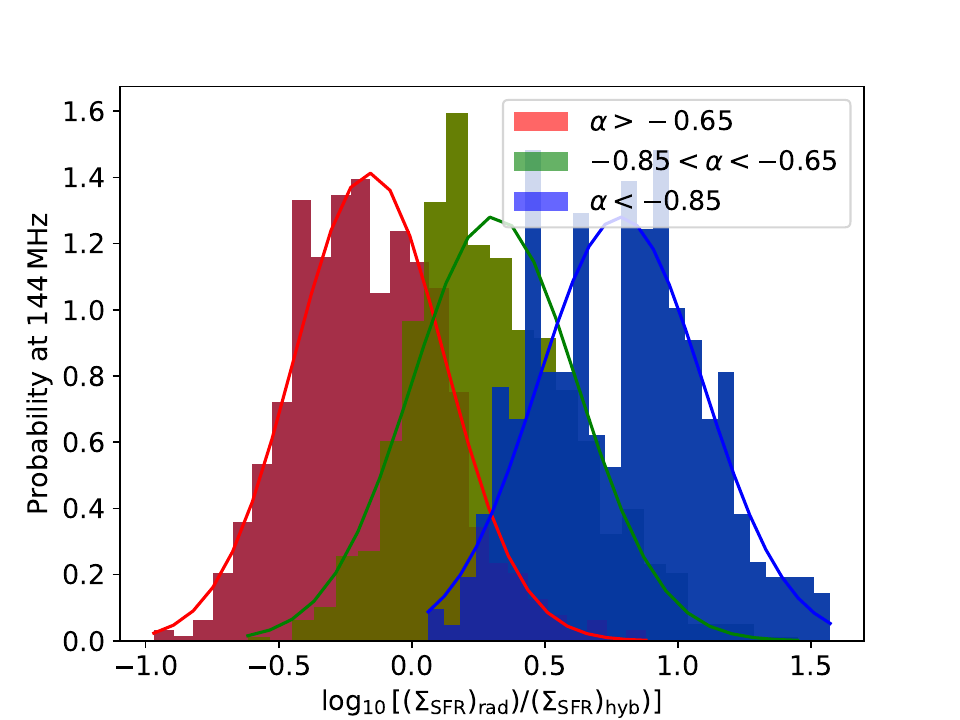}
    \end{subfigure} 
     \begin{subfigure}[t]{0.03\linewidth}
        \textbf{(b)}    
    \end{subfigure}
    \begin{subfigure}[t]{0.46\linewidth}
        \includegraphics[width=\linewidth,valign=t]{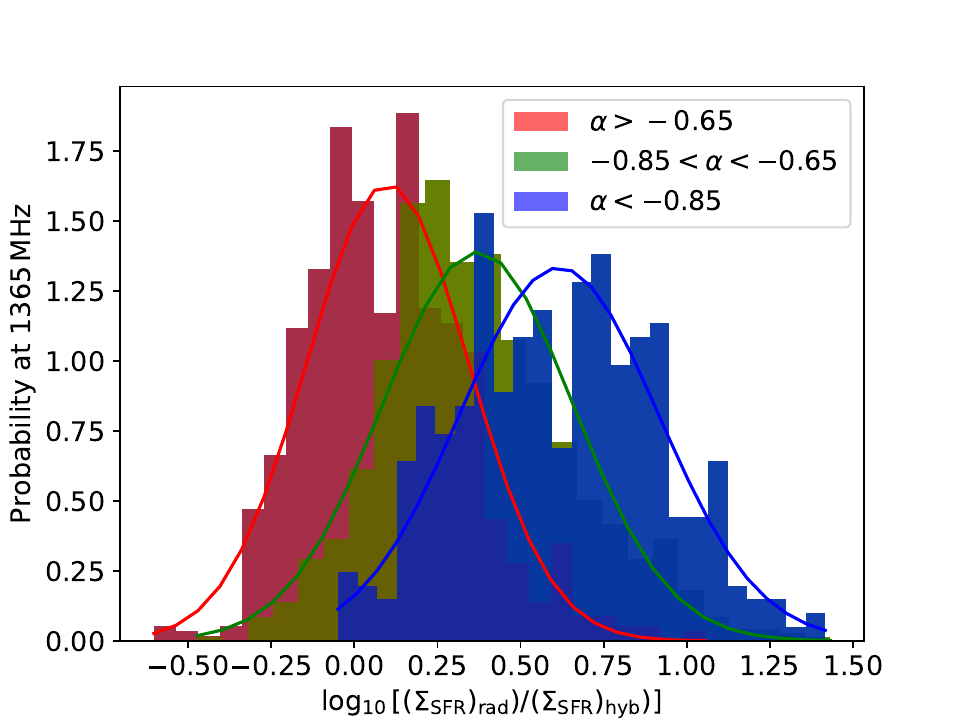}
    \end{subfigure}
    \\
     \begin{subfigure}[t]{0.03\linewidth}
        \textbf{(c)}    
    \end{subfigure}
    \begin{subfigure}[t]{0.46\linewidth}
        \includegraphics[width=\linewidth,valign=t]{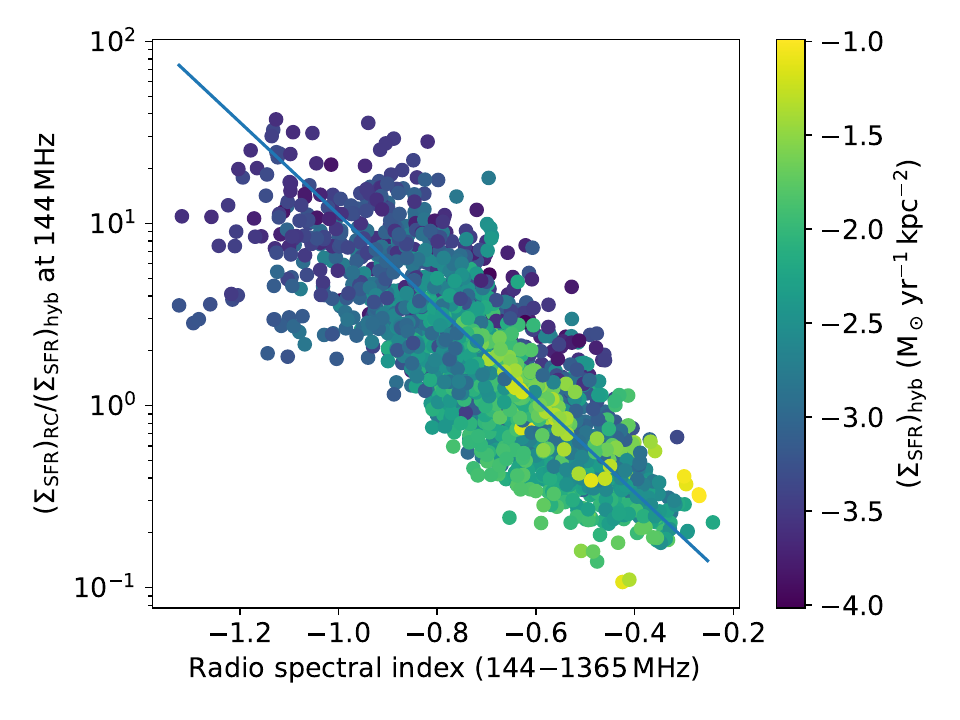}
    \end{subfigure}
     \begin{subfigure}[t]{0.03\linewidth}
        \textbf{(d)}    
    \end{subfigure}
    \begin{subfigure}[t]{0.46\linewidth}
        \includegraphics[width=\linewidth,valign=t]{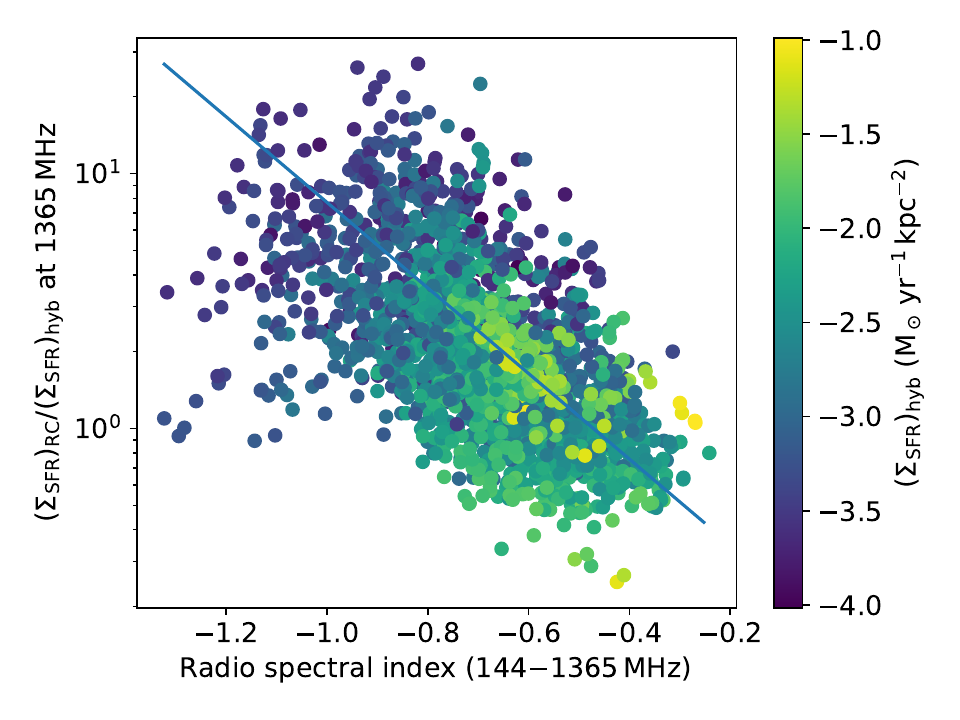}
    \end{subfigure}
    \caption{Ratio of radio to hybrid SFRD. Top panels show histograms of the ratio separately for the three radio spectral index bins at 144 MHz (panel a) and 1365 MHz (panel b). Data points representing young CREs are in red, middle-aged CREs are in green, and old CREs are in blue, respectively. Best-fitting Gaussian distributions are shown as solid lines. Bottom panels show the ratio as function of the radio spectral index between 144 and 1365\,MHz for 144\,MHz (panel c) and 1365\,MHz (panel d). Best-fitting exponential functions (Eq.\,\ref{eq:calorimetric_efficiency}) shown as solid lines.}
    \label{fig:radio_sfr_alpha}
\end{figure*}

\begin{figure*}
    \centering
         \begin{subfigure}[t]{0.03\linewidth}
        \textbf{(a)}    
    \end{subfigure}
    \begin{subfigure}[t]{0.46\linewidth}
    \includegraphics[width=\linewidth,valign=t]{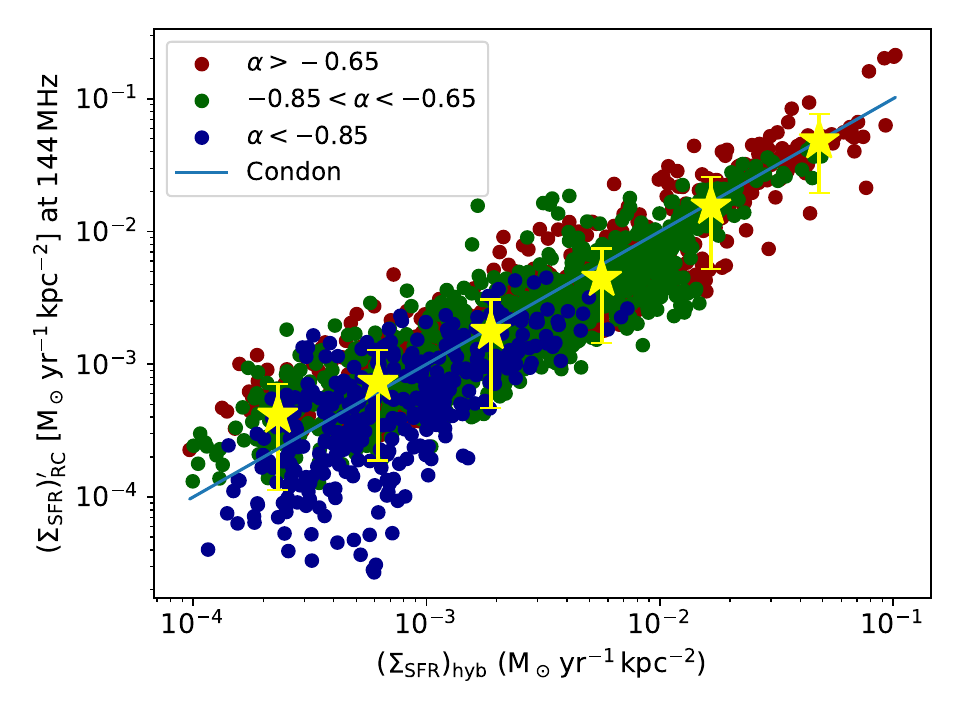}
    \end{subfigure}
             \begin{subfigure}[t]{0.03\linewidth}
        \textbf{(b)}    
    \end{subfigure}
    \begin{subfigure}[t]{0.46\linewidth}
    \includegraphics[width=\linewidth,valign=t]{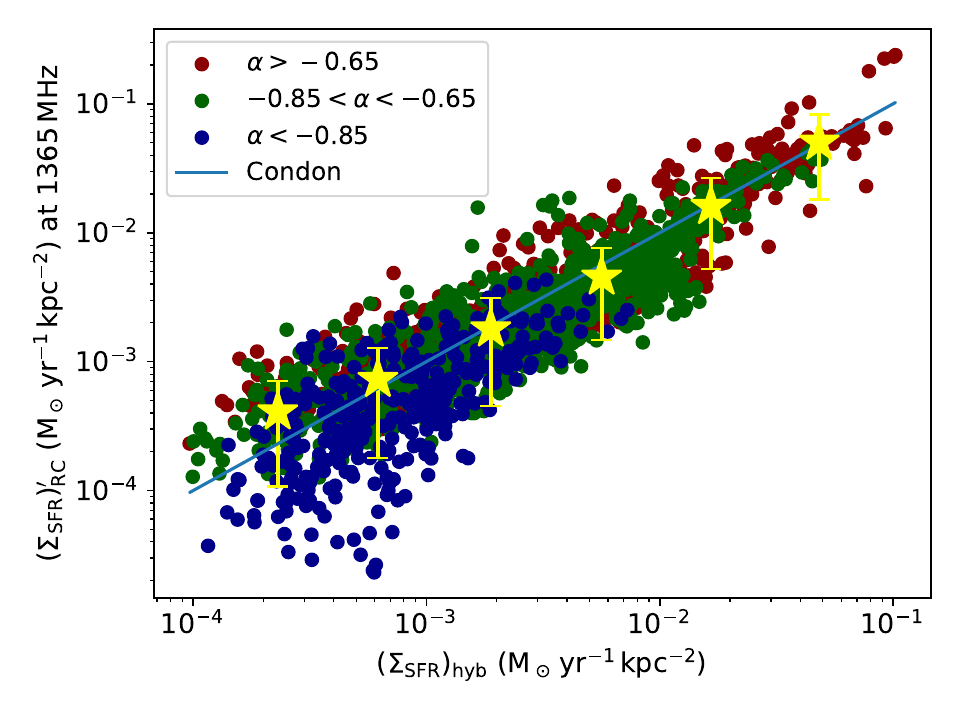}
    \end{subfigure}
    \caption{Combined radio–SFR plots corrected for the CRE calorimetric efficiency at 144 (panel a) and 1365\,MHz (panel b). We show the radio SFRD as function of the hybrid SFRD where we now corrected with the parametrisation of the calorimetric fraction as function of the radio spectral index. The 1:1 Condon relation is shown as blue line.  Data points are coloured according to the 144--1365\,MHz radio spectral index. Yellow stars show mean values. The best-fitting power-law relations are not shown as they are virtually in agreement with the Condon-relation (see Tab.\,\ref{tab:diff} for details).}
    \label{fig:radio_sfr_calor}
\end{figure*}


\begin{figure}
    \centering
    \includegraphics[width=\linewidth]{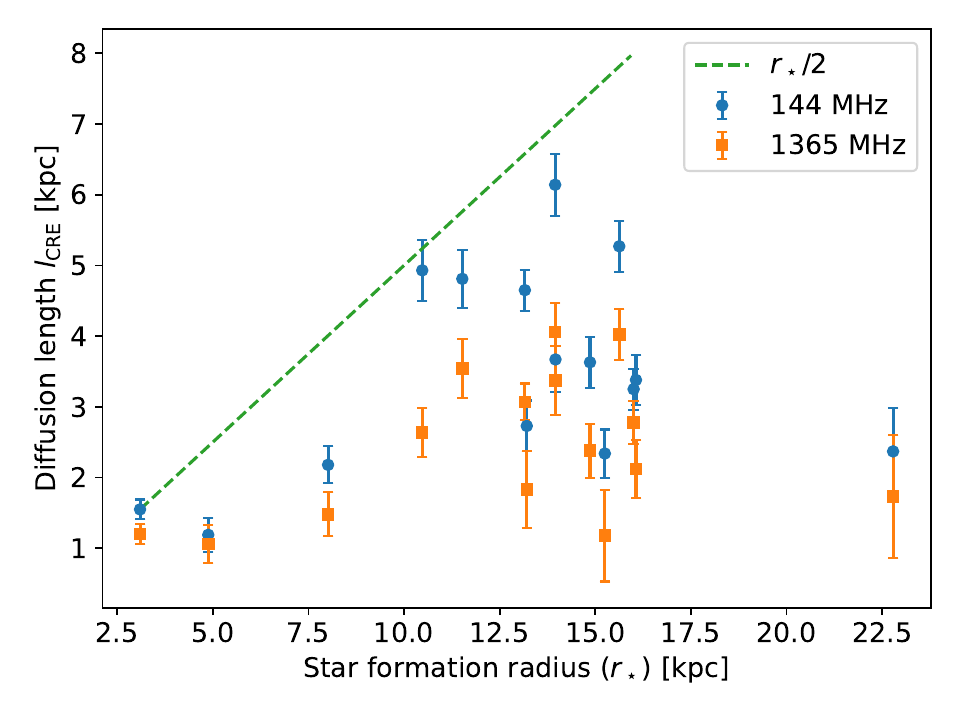}
    \caption{Diffusion length as a function of galaxy star-forming radius. Blue data points correspond to 144\,MHz and orange ones to 1365\,MHz. The dashed line shows the relation $l_{\rm CRE}=r_\star/2$.}
    \label{fig:len_radius}
\end{figure}

\begin{figure}
    \centering
    \includegraphics[width=\linewidth]{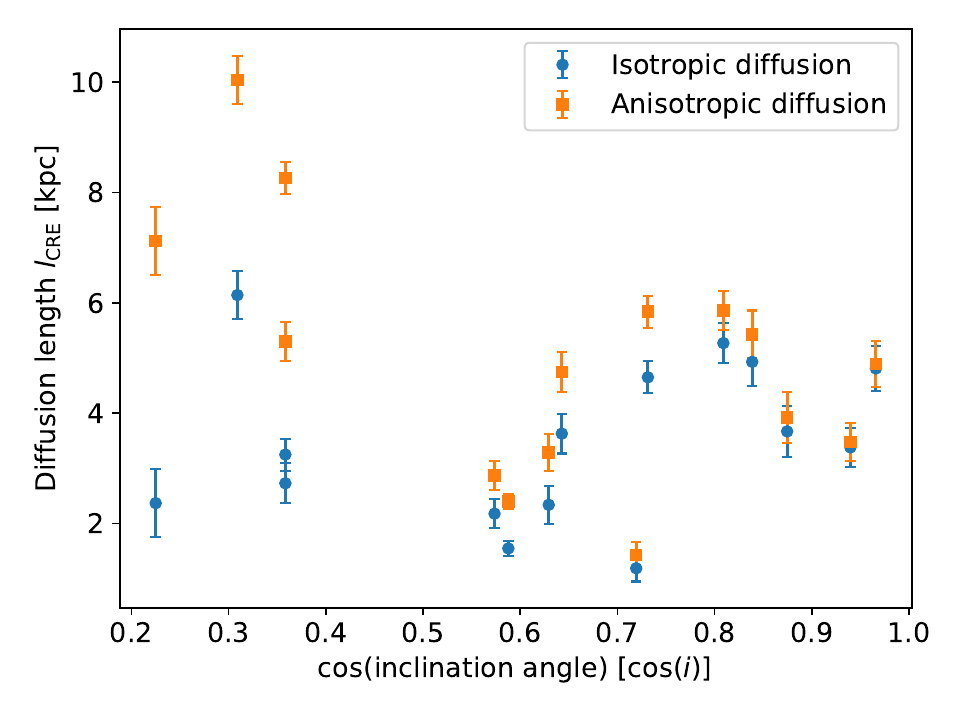}
    \caption{Diffusion length as a function of the geometric correction due to the inclination angle $\cos(i)$. Blue data points show isotropic diffusion and orange data points show anisotropic diffusion at 144\,MHz, respectively.}
    \label{fig:len_incl}
\end{figure}

\begin{figure}
    \centering
    \includegraphics[width=\linewidth]{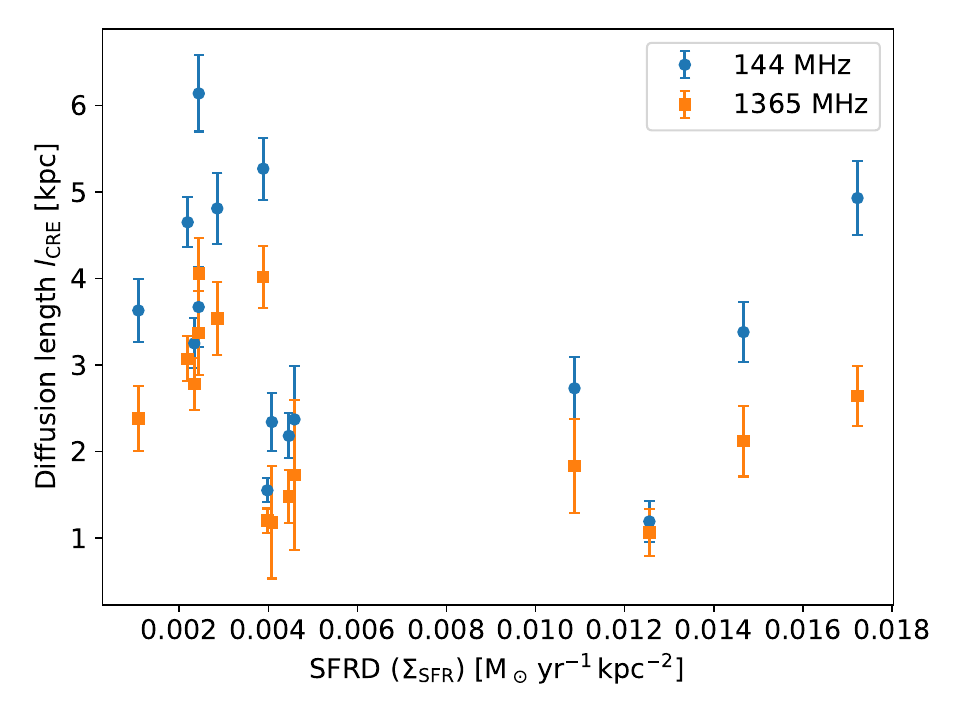}
    \caption{Diffusion length as a function of star-formation rate surface density. Data points as in Fig.\,\ref{fig:len_radius}. The two data points on the right correspond to NGC\,4254.}
    \label{fig:len_sfrd}
\end{figure}

\begin{figure}
    \centering
    \includegraphics[width=\linewidth]{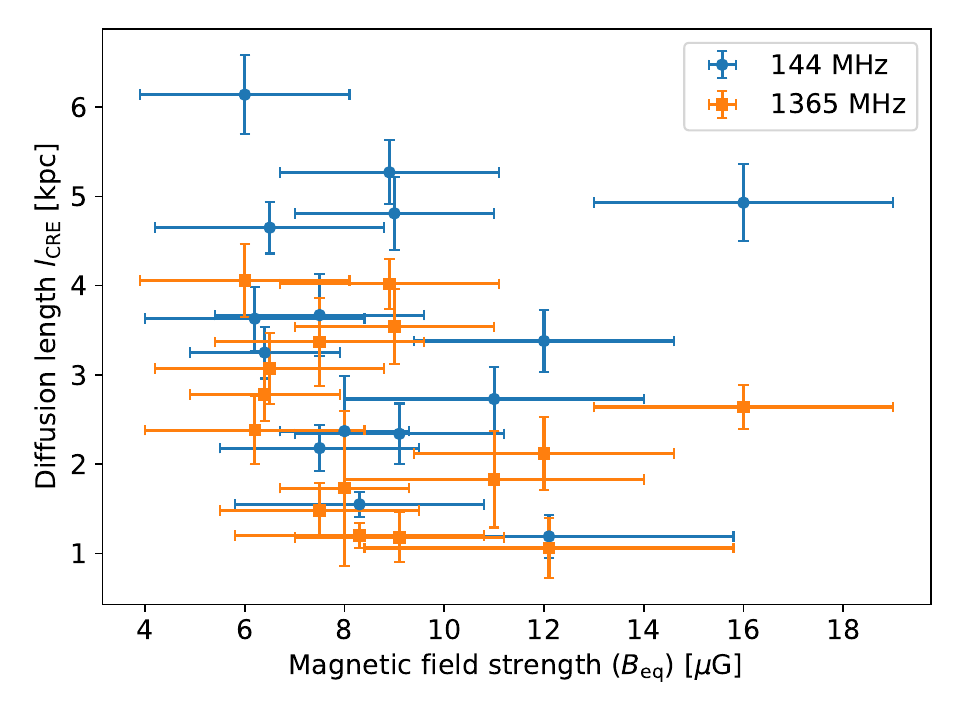}
    \caption{Diffusion length as a function of the total magnetic field strength as estimated from energy equipartition. Data points as in Fig.\,\ref{fig:len_radius}. The two data points on the right correspond to NGC\,4254.}
    \label{fig:len_bfield}
\end{figure}

\begin{figure}
    \centering
    \includegraphics[width=\linewidth]{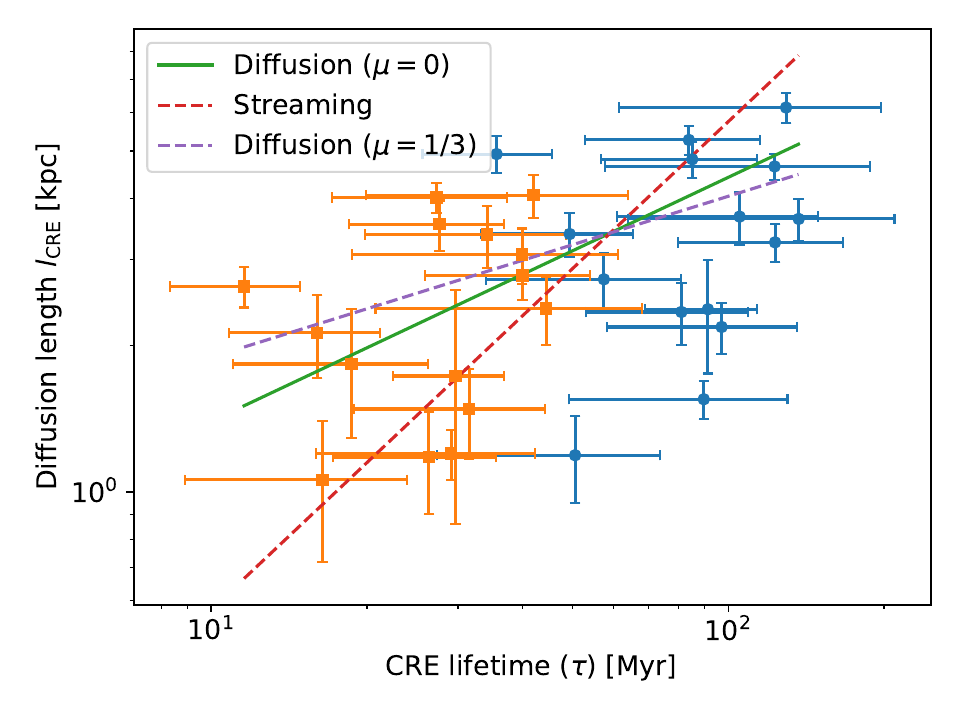}
    \caption{Diffusion length as a function of CRE lifetime. Data points as in Fig.\,\ref{fig:len_radius}. Cosmic-ray transport models are shown for comparison. The green solid line shows energy-independent diffusion ($\mu=0$), the dashed red lines shows streaming, and the purple dashed line energy-dependent diffusion ($\mu=1/3$). The diffusion coefficient is parametrised as $D\propto E^\mu$.}
    \label{fig:len_tau}
\end{figure}

\section{Results}
\label{s:results}
\subsection{Spatially resolved radio continuum--star formation rate relation}
\label{ss:radio-sfr}

The spatially resolved analysis shows that the data points follow a sublinear radio--SFR relation for both frequencies 144 and 1365\,MHz (Figs.\,\ref{fig:radio_sfr}a and b). The slope of the relation is flatter for lower frequencies which is expected because we see an older and therefore lower energy population of CREs, which had more time to diffuse. We divide the points into three groups based on their radio spectral indices: young CREs ($-0.65 < \alpha < -0.20 $; $\lesssim$30\,Myr), medium age CREs ($-0.85 < \alpha < -0.65 $; 30--100\,Myr) and old CREs ($\alpha<-0.85$; $\gtrsim$100\,Myr). The values for the spectral indices are typical values found in the spiral arms, inter arm regions and outskirts of galaxies, respectively.

Focusing just on the young CREs (Figs.\,\ref{fig:radio_sfr}c and d) we see the slopes of their radio--SFR relations are a lot closer to unity and there is no significant difference between the frequencies. This behaviour is explained by the fact that the young CREs have energy spectra very close to their injection spectra. They have not propagated far from their sources and thus have not lost a lot of energy. This implies their spatial distribution should more closely resemble that of the star formation and they should therefore follow the Condon-relation more closely. The results are shown in Table\,\ref{tab:diff}.

\begin{table*}
	\centering
	\caption{Radio--SFR relation and CRE transport lengths in galaxies of the sample.}
	\label{tab:diff}
        \begin{tabular}{l cccc ccc} 
          \hline\hline
          
          Galaxy & $a_1$ & $b_1$ & $a_1$ & $b_2$ & $l_1$  & $l_2$ & $l_1/l_2$   \\ 
   & & & & & (kpc) & (kpc) \\
       (1) & (2)  & (3) & (3) & (4) & (5) &  (6) & (7)\\ \hline
    NGC 0628 & $0.41\pm 0.02$ & $-1.73\pm 0.05$ & $0.58\pm 0.02$ & $-1.07\pm 0.05$ & $4.81\pm 0.41$ & $3.54\pm 0.42$ & $1.36\pm 0.20$ \\
    NGC 0925 & $0.50\pm 0.03$ & $-1.69\pm 0.08$ & $0.69\pm 0.02$ & $-0.83\pm 0.07$ & $3.63\pm 0.36$ & $2.38\pm 0.38$ & $1.53\pm 0.29$ \\ 
    NGC 2403 & $0.58\pm 0.02$ & $-1.37\pm 0.06$ & $0.76\pm 0.02$ & $-0.64\pm 0.06$ & $2.18\pm 0.26$ & $1.48\pm 0.31$ & $1.47\pm 0.36$ \\ 
    NGC 2841 & $0.44\pm 0.02$ & $-1.17\pm 0.06$ & $0.54\pm 0.02$ & $-0.84\pm 0.06$ & $3.25\pm 0.29$ & $2.78\pm 0.30$ & $1.17\pm 0.16$ \\
    NGC 2903 & $0.62\pm 0.02$ & $-0.78\pm 0.04$ & $0.83\pm 0.02$ & $-0.05\pm 0.05$ & $2.73\pm 0.36$ & $1.83\pm 0.54$ & $1.49\pm 0.48$ \\
    NGC 2976 & $0.43\pm 0.03$ & $-1.74\pm 0.10$ & $0.58\pm 0.03$ & $-1.05\pm 0.10$ & $1.55\pm 0.14$ & $1.20\pm 0.14$ & $1.29\pm 0.19$ \\
    NGC 3184 & $0.58\pm 0.02$ & $-1.23\pm 0.06$ & $0.63\pm 0.02$ & $-0.92\pm 0.05$ & $3.92\pm 0.46$ & $3.61\pm 0.49$ & $1.09\pm 0.21$ \\ 
    NGC 3198 & $0.31\pm 0.02$ & $-1.91\pm 0.05$ & $0.51\pm 0.02$ & $-1.26\pm 0.05$ & $6.13\pm 0.44$ & $4.06\pm 0.41$ & $1.51\pm 0.19$ \\
    NGC 3938 & $0.26\pm 0.03$ & $-1.67\pm 0.06$ & $0.44\pm 0.03$ & $-1.16\pm 0.06$ & $5.27\pm 0.36$ & $4.02\pm 0.28$ & $1.31\pm 0.15$ \\
    NGC 4254 & $0.43\pm 0.01$ & $-0.75\pm 0.03$ & $0.62\pm 0.01$ & $-0.33\pm 0.03$ & $4.93\pm 0.43$ & $2.64\pm 0.25$ & $1.87\pm 0.30$ \\
    NGC 4725 & $0.20\pm 0.03$ & $-2.43\pm 0.09$ & $0.42\pm 0.03$ & $-1.56\pm 0.09$ & $4.65\pm 0.29$ & $3.07\pm 0.40$ & $1.51\pm 0.16$ \\
    NGC 4736 & $0.75\pm 0.02$ & $-0.63\pm 0.05$ & $0.80\pm 0.02$ & $-0.23\pm 0.06$ & $1.19\pm 0.24$ & $1.06\pm 0.34$ & $1.12\pm 0.36$ \\
    NGC 5055 & $0.66\pm 0.01$ & $-0.52\pm 0.04$ & $0.91\pm 0.02$ & $0.04\pm 0.05$  & $2.34\pm 0.34$ & $1.18\pm 0.28$ & $1.98\pm 1.12$ \\ 
    NGC 5194 & $0.52\pm 0.02$ & $-0.70\pm 0.04$ & $0.74\pm 0.02$ & $-0.13\pm 0.04$ & $3.38\pm 0.35$ & $2.12\pm 0.41$ & $1.59\pm 0.35$ \\ 
    NGC 7331 & $0.81\pm 0.03$ & $-0.01\pm 0.07$ & $0.90\pm 0.02$ & $0.24\pm 0.06$  & $2.37\pm 0.62$ & $1.73\pm 0.87$ & $1.37\pm 0.78$ \\
    Sample   & $0.86\pm 0.02$ & $-0.14\pm 0.01$ & $0.92\pm 0.01$ & $0.08\pm 0.01$  & $3.5\pm 1.4$   & $2.4\pm 1.0$   & $1.44\pm 0.24$ \\
    Young CRE$\tablefootmark{a}$  & $1.06\pm 0.02$ & $-0.02\pm 0.01$  & $1.03\pm 0.02$ & $0.15\pm 0.01$ & N/A            & N/A            & N/A            \\
    Sample-cor$\tablefootmark{b}$ & $1.02\pm 0.01$ & $0.05\pm 0.02$  & $1.00\pm 0.01$ & $0.00\pm 0.02$  & N/A            & N/A            & N/A            \\
    \hline
    \end{tabular}
    \tablefoot{Columns (2+4) $a_1$ and $a_2$ are the slopes at 144 and 1365\,MHz, respectively; (3+5) $b_1$ and $b_2$ are the the $y$-axis intercepts at 144 and 1365\,MHz, respectively (see Eq.\,\ref{eq:loglog_fit} for a definition of $a$ and $b$); (6+7) $l_1$ and $l_2$ are the CRE transport lengths at 144 and 1365\,MHz, respectively; (8) is the ratio $l_1/l_2$. \tablefoottext{a}{radio--SFR relation for young CREs (see Sect.\,\ref{ss:radio-sfr})}; \tablefoottext{b}{radio--SFR relation for sample corrected for the CRE calorimetric efficiency (see Sect.\,\ref{ss:calorimetric_efficiency})}
    }
\end{table*}

\subsection{Influence of the calorimetric efficiency}
\label{ss:calorimetric_efficiency}

In this section we investigate the influence of the calorimetric efficiency on the radio--SFR relation. The calorimetric efficiency  is defined as the relative energy loss of the CREs due to synchrotron radiation \citep{pfrommer_22a,heesen_23a}. Hence we assume
\begin{equation}
    (\Sigma_{\rm SFR})_{\rm RC} = \eta (\Sigma_{\rm SFR})_{\rm hyb}.
    \label{eq:calorimetric_sfr}
\end{equation}
 In Figs.\,\ref{fig:radio_sfr_alpha}(a) and (b) we show a probability density distribution of $\log_{10}[(\Sigma_{\rm SFR})_{\rm RC}/(\Sigma_{\rm SFR})_{\rm hyb}]$, the logarithmic ratio of radio to hybrid SFR surface density, separately for our three radio spectral index bins. While for young CREs the distribution is centred around $-0.3$--$0$, equating to ratios of $0.5$--1, so that the radio and hybrid approximately \sfrd\ agree, for middle aged and old CREs the distribution is markedly different. For old CREs the distribution is centred around $0.7$--$0.9$, equating to ratios of 5--8 which is larger than one, so that the radio \sfrd\ is significantly higher than the hybrid \sfrd. This effect is even more pronounced for 144\,MHz when compared with 1365\,MHz. 

In the next step we quantified this dependence further by studying the ratio of radio to hybrid \sfrd\ as function of radio spectral index. This is shown in Figs.\,\ref{fig:radio_sfr_alpha}(c) and (d).  We find a clear correlation between the ratio and the radio spectral index. Such a behaviour can be explained by a varying calorimetric fraction of the CRE, so that the calorimetric efficiency can be parametrised with the radio spectral index
\begin{eqnarray}
    \nonumber
    \eta(\alpha)_{\rm 144\,\rm MHz} & = &\exp[(-5.85\pm 0.08) \alpha-(3.43\pm 0.06)]\\
    \eta(\alpha)_{\rm 1365\,\rm MHz} & = &\exp[(-3.86\pm 0.08) \alpha-(1.82\pm 0.06)].
    \label{eq:calorimetric_efficiency}
\end{eqnarray}
Recall that if the CREs lose a large fraction of their energy, the radio spectral index becomes $\alpha = \alpha_{\rm inj}-0.5$, where $\alpha_{\rm inj}$ is the injection spectral index. In contrast, if the CREs lose almost no energy via synchrotron radiation losses, then we expect a radio spectral index similar to the injection spectral index. Thus we expect the calorimetric efficiency to be a function of radio spectral index as we have now confirmed.

We can now attempt to correct for the calorimetric efficiency by parametrising it with the radio spectral index and then divide the radio \sfrd\ value by this (Eq.\,\ref{eq:calorimetric_sfr}). The corrected radio \sfrd is hence
\begin{equation}
(\Sigma_{\rm SFR})_{\rm RC}^\prime = (\Sigma_{\rm SFR})_{\rm RC} / \eta(\alpha)
\label{eq:calorimetric_sfrd}
\end{equation}
In Fig.\,\ref{fig:radio_sfr_calor}, we show the resulting radio--SFR relation. We can now see that this correction is able to restore a linear relation in close agreement with the Condon-relation. We also show mean values binned in hybrid \sfrd. Best-fitting relations are presented in Table\,\ref{tab:diff}.

\subsection{Cosmic-ray transport lengths}

The principal results from the smoothing experiment are the CRE transport lengths (Table\,\ref{tab:diff}). At 144\,MHz we find a mean CRE transport length of $\langle l_1\rangle=3.5\pm 1.4$\,kpc and at 1365\,MHz we find $\langle l_2\rangle = 2.4\pm 1.0$\,kpc. First, we would like to rule out that the size of our galaxies has any systematic effect when measuring the CRE transport lengths. In Fig.\,\ref{fig:len_radius} we plot the transport lengths as a function of the star-formation radii. We find no significant correlation between $l_{\text{CRE}}$ and $r_\star$ with a Spearman rank correlation coefficient of $\rho_s=0.22$. However, we find that the diffusion length is limited by the size of the galaxy with $l_{\rm CRE}<r_\star/2$. The diffusion length cannot be larger than the size of the galaxy. If the galaxies are too small, it may be impossible to detect older CREs effectively. Their lifetimes are so large and their transport lengths so long that they escape the galaxy before losing a lot of their energy. Outside the galaxy the magnetic field strength is weak and therefore these electrons become undetectable. Secondly, it is hard to accurately measure values for $l_{\text{CRE}}$ that are smaller than the resolution of our maps, hence $\approx$1\,kpc is the lower limit we can detect with this method.

Next we investigated the possible influence of the inclination angle on the CRE transport length. We show the resulting diffusion lengths as a function of the geometric correction factor (see Eq.\,\ref{eq:inclination}) in Fig.\,\ref{fig:len_incl}. Again, we would expect no influence as the transport length of cosmic rays in the disc should be independent of inclination angle. For the isotropic diffusion kernel this is indeed the case as the CRE transport length at 144\,MHz is almost equal whether measured in galaxies for low ($\cos(i)>0.5$) or high ($\cos(i)<0.5$) inclination angles. On the other hand for the anisotropic diffusion kernel we find $l_{\rm CRE}=4.0\pm 0.4$\,kpc at low and $l_{\rm CRE}=7.7\pm 0.9$\,kpc at high inclination angles. Hence, we find that the anisotropic diffusion clearly overestimates the CRE transport length in the latter galaxies. This justifies our choice of not correcting for the inclination angle because the cosmic rays diffuse in isotropic fashion so that there is no strong preferential diffusion in the disc plane of the galaxies \citep[see also][]{murphy_08a}.

\subsection{Cosmic-ray electron transport}
In this section we investigate the properties of CRE transport in more detail. We calculate the electron lifetime  for which we use estimates of the total magnetic field strength. We use the measurements of \citet{heesen_23a} who used the revised equipartition formula by \citet{beck_05a} and the 144-MHz maps together with the low-frequency radio spectral index. For the additional galaxies not included in the sample of \citet{heesen_23a} we used values from the literature (see Table\,\ref{tab:sample} for details). In order to take inverse Compton losses into account we calculated the radiation energy density $U_{\text{rad}}$. Here we follow the approach used in \citet{heesen_18a} combining the CMB energy density, which can be measured directly, and the total radiation energy density which is estimated using the total infrared luminosity.

Then we investigated diffusion lengths in relation to SFR surface densities (Fig.\,\ref{fig:len_sfrd}). \citet{murphy_08a} found that high SFR densities lead to lower CRE transport lengths. The reasoning is that more SFR  density means more SNe which lead to more shocks and amplify the $B$ field in a galaxy. That should decrease the average transport length because synchrotron losses increase with magnetic field strength. We find only tentative indication of such a correlation ($\rho_s=-0.42$); however, if we leave out NGC~4254 which is affected, both, by ram pressure and strong gravitational perturbation \citep{vollmer_05a,chyzy_07a,duc_08a,boselli_18a}, the correlation improves to $\rho_s=-0.57$. We then explored the correlation with the equipartition magnetic field strength in Fig.\,\ref{fig:len_bfield} and found a weaker correlation with $\rho_s=-0.45$ (and $\rho_s=-0.32$ including NGC~4254). This could indicate that it is not only the magnetic field strength and hence the CRE lifetime that determines the transport length.

We can write energy $E$ and lifetime $\tau$ of the CREs as \citep{heesen_19a}:
\begin{align}
\nonumber
    E({\rm GeV}) &= \sqrt{  \left(\frac{\nu }{16.1\,{\rm MHz}}\right) \left( \frac{B_{\rm eq}} {{\muup\rm G}}\right)^{-1} } \\
    \tau &= 8.352\times 10^9 \left( \frac{E}{\rm GeV}\right)^{-1} \left( \frac{B_{\rm eq}}{\rm \muup G}\right)^{-2} \left( 1 + \frac{U_{\text{rad}}}{U_{\text{B}}} \right)^{-1}~\rm yr \,.
\end{align}
We can use the relation between CRE lifetime and transport length to discriminate between the different possible models of CRE transport. The models predict different scaling of CRE transport length $l_{\rm CRE}$ and lifetime $\tau$ with frequency $\nu$. For diffusion we expect $l_{\text{diff}} \propto \nu^{-1/4}$ and for advection $l_{\text{adv}} \propto \nu^{-1/2}$ \citep{vollmer_20a}. Another  possibility we  consider is an energy-dependent diffusion coefficient $D(E)\propto E^{1/2}$ which would cause a scaling proportional to $\nu^{-1/8}$. In general, for a diffusion coefficient $D\propto E^\mu$ we expect $l_{\rm diff}\propto \nu^{(\mu-1)/4}$ \citep[e.g.][]{heesen_21a}. Since $\tau \propto \nu^{-0.5}$ we can plot diffusion length as a function of lifetime in a double logarithmic plot and the points should follow a linear relation with the slope twice as high as when plotted versus frequency (Fig. \ref{fig:len_tau}). The data points show a relatively large scatter, although the correlation found is the most significant yet ($\rho_{\rm s}=0.61$) when NGC\,4254 is omitted. While our data are in agreement with energy-independent diffusion ($\mu=0$) we cannot rule out a modest energy-dependence of the diffusion coefficient such as $\mu=1/3$; similarly, we cannot rule out cosmic-ray streaming either, even though the slope appears to be steeper than what is observed.

Another way to look at different cosmic-ray transport models is to directly compare the diffusion lengths at two different frequencies. We compute $l_1/l_2$ for our sample, which is the ratio of 144 to 1365\,MHz CRE transport lengths. The results are shown in Table\,\ref{tab:diff}. The mean value is $\langle l_1/l_2\rangle = 1.44\pm 0.24$. This value lies between the theoretical value for energy-dependent diffusion of $1.32$ and energy-independent diffusion of $1.75$. However, there are reasons why energy-independent diffusion is more likely. If we ignore CRE escape completely, our results would suggest some galaxies have energy-dependent diffusion while others have energy-independent diffusion. In that case one would have to find an explanation why different galaxies have different transport mechanisms. It is more likely that those galaxies that have low ratios have faster CRE escape than those with high ratios. There are only two galaxies that have ratios in excess of what is expected for energy-dependent diffusion. This could be explained by cosmic-ray streaming or advection. These are NGC~4254 and 5055. In NGC~4254 this is expected as it has a radio tail where cosmic rays are advected or stream down the pressure gradient \citep{chyzy_07a}. NGC~5055 is currently undergoing a minor merger with a disturbed outer disc \citep{chonis_11a}. Large-scale gas motions would lead to CRE advection explaining our results. We therefore conclude that our results make energy-independent CRE diffusion most likely.

\begin{figure}
    \centering
    \includegraphics[width=\linewidth]{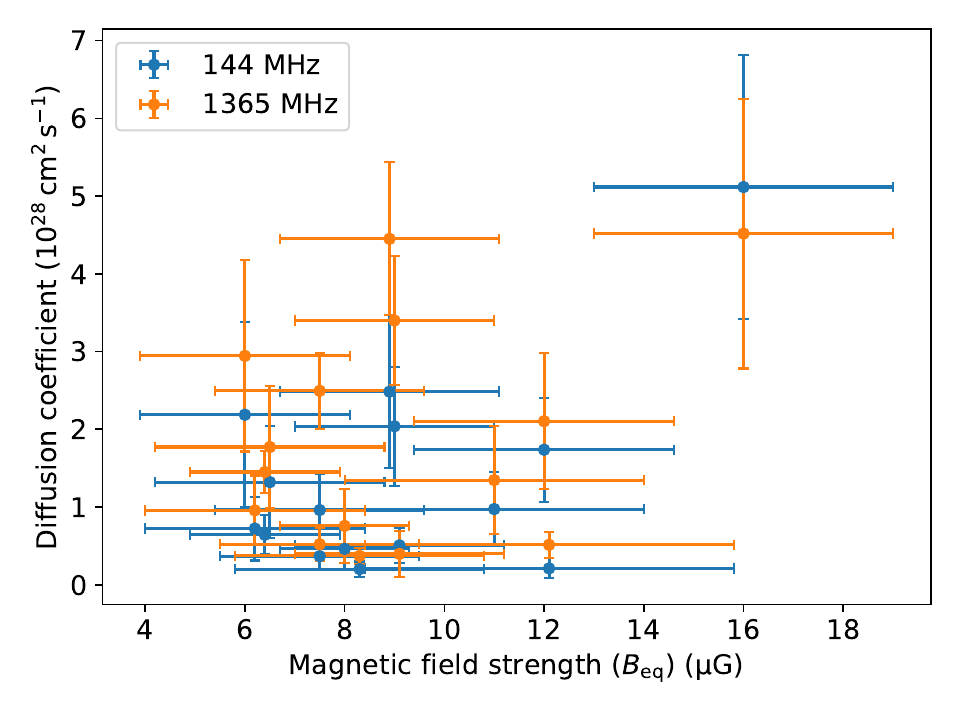}
    \caption{Isotropic diffusion coefficient as function of total magnetic field strength as estimated from energy equipartition in microgauss. Blue data points correspond to 144\,MHz and orange ones to 1365\,MHz. The data points on the right correspond to NGC~4254.}
    \label{fig:diff_bfield}
\end{figure}

\subsection{Cosmic-ray diffusion and magnetic fields}

Now that we have established that cosmic-ray diffusion can describe our data best, we investigate the influence of magnetic fields on CRE diffusion. To calculate the diffusion coefficient $D$ we use the relation between lifetime $\tau$ and diffusion length $l_{\text{CRE}}$ for isotropic 3D diffusion:
\begin{equation}
    D = \frac{l_{\text{CRE}} ^2}{4\tau}.
\end{equation}
The mean diffusion coefficient is $\langle D\rangle = (1.07 \pm 0.16)\times 10^{28}$\,\udif.

In Fig.\,\ref{fig:diff_bfield} we plot the diffusion coefficient as a function of the total magnetic field strength. There is a high variance in the values for the diffusion coefficients with no clear trend visible. The highest values are found in NGC~4254, where the radio tail will raise the values. The theory of 1D CRE diffusion in a disc suggests that the diffusion coefficient scales as \citep{shalchi_09a}
\begin{align}
    D \propto \left( \frac{B_{\text{ord}}}{B_{\text{turb}}} \right)^2  B_{\text{turb}} ^{-1/3},
\end{align}
where $B_{\text{ord}}$ is the ordered magnetic field component  and $B_{\text{turb}}$ is the turbulent magnetic field component. That means we should not expect to see a direct correlation with the total magnetic field strength $B_{\rm eq}$ or perpendicular magnetic field strength $B_{\perp}$. We cannot rule out any possible weak dependence of $D$ on $B_{\rm eq}$. The high variance in the data and the small range of values for $B_{\rm eq}$ would make it hard to detect.

The implied strong correlation of $D$ with $(B_{\text{ord}}/B_{\text{turb}})^2$ can be tested: the total perpendicular magnetic strength can be decomposed $B_{\rm eq}  ^2 = B_{\text{ord}} ^2 + B_{\text{turb}} ^2$. In principle it is possible to measure two of those three: $B_{\rm eq}$ is proportional to  the total radio intensity and $B_{\text{ord}}$ is proportional to the polarized radio intensity \citep[e.g.][]{tabatabaei_16a}. When linearly polarized radio waves traverse a plasma, they undergo Faraday depolarization. This process occurs, both, in the ISM and the Earths' atmosphere and is proportional to $\nu^{-2}$. In practice, at low frequencies the radio continuum radiation is almost completely depolarized, making it very hard to directly probe the components of the magnetic field (see Tab.\,\ref{tab:sample}).

\begin{figure}
    \centering
    \includegraphics[width=\linewidth]{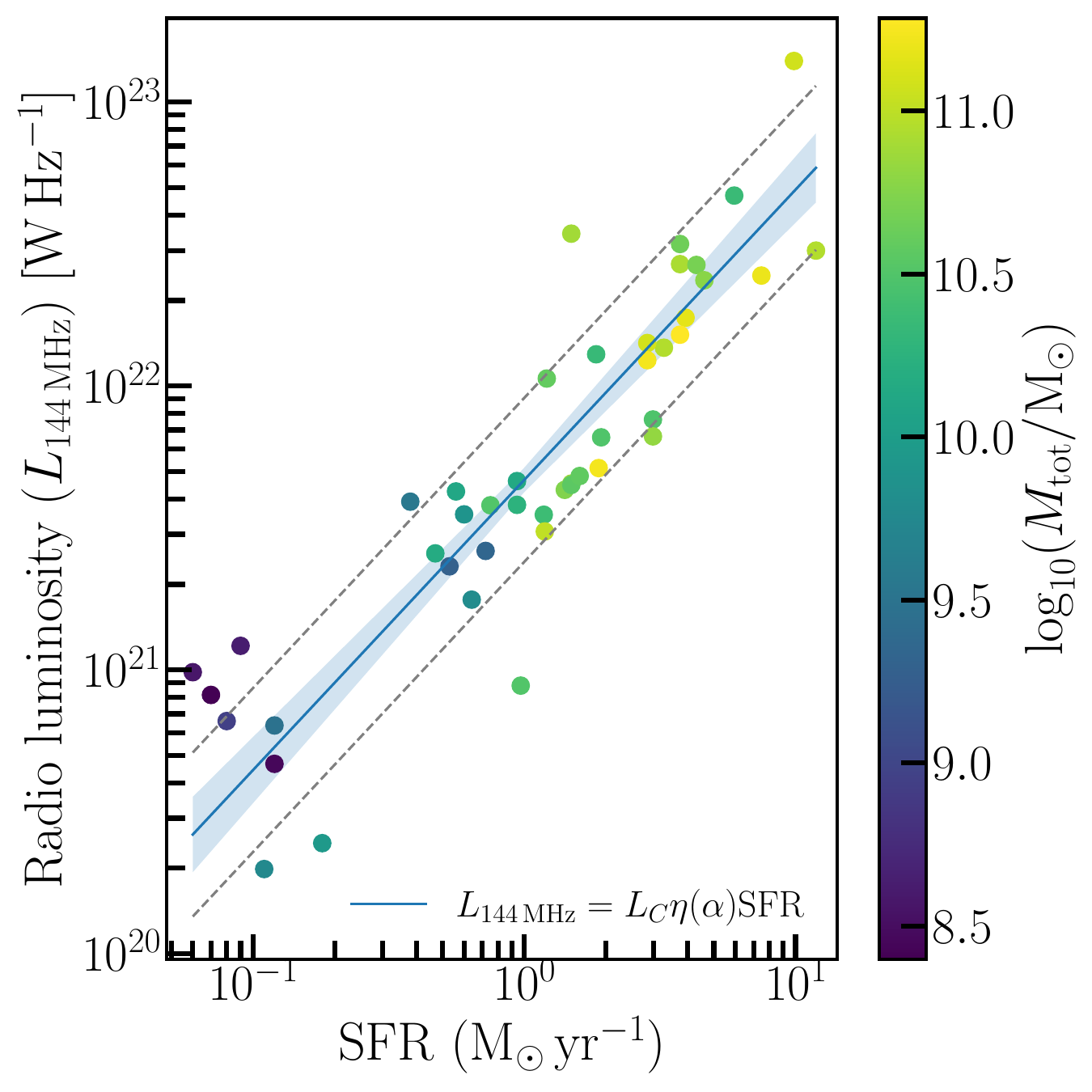}
    \caption{Integrated radio--SFR relation but corrected for calorimetric efficiency parametrised by the radio spectral index. Blue line and shades areas show best-fitting relartion with uncertainties. Dashed lines show 1$\sigma$ scatter of data points.}
    \label{fig:integrated_radio_sfr}
\end{figure}
\section{Discussion}
\label{sec:discussion}
\subsection{Radio continuum--star formation rate relation}
\subsubsection{Semicalorimetric radio--SFR relation}
In this work we calculated the spatially resolved radio--SFR relation in 15 nearby galaxies at a resolution of approximately (1.2\,kpc)$^2$. We used two different observing frequencies and found that for all of them the slopes of the radio--SFR relation for the full sample were smaller than one. We find $a_1=0.86\pm 0.02$ at 144\,MHz and $a_2=0.92\pm 0.01$ at 1365\,MHz. These results are in agreement with CRE transport that predict CREs move away from their places of origin. As a result, spots with high SFR surface densities will appear radio dim and places with low SFR surface densities will appear radio bright. That also explains the fact that the slope becomes flatter at lower frequencies. At lower frequencies we observe older CREs with a more smoothed out distribution. If we focus just on the young CREs (areas with spectral indices $\alpha >-0.65$), the radio--SFR relation becomes almost linear with no apparent correlation with the frequency. We find $a_1=1.06\pm 0.02$ at 144\,MHz and $a_2=1.03\pm 0.02$ at 1365\,MHz. Those relations are expected to be steeper than the ones for all CREs since the young CREs have not had time to move very far from their origins. Hence, their distribution should be (almost) identical to the distribution of star formation. The fact that they follow a nearly linear relation could be a hint that there is some kind of local calorimetry. The CREs are somehow confined in the spiral arms and lose all their energy there. 

In contrast, the global radio--SFR relation holds over five orders of magnitude and is super-linear with most slopes reported between $1.1$ and $1.4$ where lower frequencies tend to have higher slopes \citep{li_16a}. The super-linear slope cannot be explained with a pure calorimeter model which is why \citet{niklas_97a} proposed an equipartition model. First, it  requires the energy stored in cosmic rays and the magnetic field to be equal. Second, it proposes a feedback mechanism to explain the superlinearity: higher SFR leads to more gas turbulence which in turn amplifies the magnetic field. The increased magnetic field strength is then responsible for the increased synchrotron radiation. Their proposed relation is  \citep[see also][]{schleicher_13a}:
\begin{align}
 \text{SFR} \propto  B_{\rm eq} ^{1/(3-\alpha)}\,, \label{eq:niklas}
\end{align}
where $\alpha \approx -0.6$ is the CRE radio spectral index at injection. 

There are however some recent results that seem to point more towards a semi-calorimetric model, where the escape of CREs play also a role.
A large study by \citet{smith_21a} found an (almost) linear radio--SFR relation at low radio frequencies ($1.041 \pm 0.007$) in line with earlier results from \citet{gurkan_18a} who used a similar methodology with a smaller sample. \citet{gurkan_18a} and \citet{smith_21a} also both found a mass dependence of their radio--SFR relation. At a constant SFR, the more massive a galaxy is, the more luminous it is in the radio continuum. It would be difficult to explain this behaviour with a calorimetric model alone. That is why the authors suggest that  mass dependent CRE escape in these galaxies could be the reason for their findings. \citet{chyzy_18a} investigated integrated radio spectral indices in star-forming galaxies and found that they were flatter than their injection indices $\alpha \approx -0.57$. When the sample was restricted to frequencies above 1.5\,GHz the radio spectra became steeper with $\alpha \approx -0.77$. If galaxies were perfect calorimeters we should see an integrated radio spectral index of $\alpha \approx -1.1$. Since that is not the case there seems to be significant escape of CREs, in particular below 1\,GHz.

\subsubsection{Updated radio--SFR calibration}
Hence, we now probe the effect of applying the calorimetric correction (Sect.\,\ref{ss:calorimetric_efficiency}) on the integrated radio--SFR relation. We can write the integrated radio--SFR relation as
\begin{equation}
    L_{\rm 144\,MHz}=L_{\rm C}\eta(\alpha){\rm SFR}^c,
    \label{eq:integrated_radio_sfr}
\end{equation}
where we have used the calorimetric efficiency parametrised with the radio spectral index as given by Eq.\,\eqref{eq:calorimetric_efficiency}.
We used the 45 galaxies of \citet{heesen_22a} with the result shown in Fig.\,\ref{fig:integrated_radio_sfr}. The best-fitting relation has a slope of $c=1.02\pm 0.07$ with $L_{\rm C}=10^{21.67\pm 0.04}\,\rm W\,Hz^{-1}$. Notably, the relation is now almost linear and in excellent agreement with the slope of the spatially resolved relation. The constant $L_{\rm C}$ is in excellent agreement with what is predicted by the spatially resolved relation (Equation~\ref{eq:condon_final}). Combined with our two measurements using the spatially resolved observations, our best estimate for the slope of the radio--SFR relation is $1.01\pm 0.02$.

This means that we have for the first time identified a universal radio-SFR relation that can both be applied to integrated as well as spatially resolved measurements. This means that the observed variation in the slope of the radio--SFR relation can be explained by the influence of the CRE calorimetric efficiency. This was already proposed by \citet{smith_21a} who speculated that a higher mass leads to higher efficiency. This was corroborated by \citet{heesen_22a} who found a correlation between radio spectral index and total mass. A similar relation is found for galaxies in the Virgo cluster in the LOFAR survey by \citet{edler_23a} and Edler et al. in preparation. The finding that this can also explain the local radio--SFR relation strongly corroborates this interpretation. We can now provide updated calibrations of the radio--SFR relation that absorbs the calorimetric efficiency into the radio continuum luminosity. To this end we combine the radio spectral index--SFR relation \citep{heesen_22a} with Eq.\,\eqref{eq:calorimetric_efficiency} to obtain
\begin{eqnarray}
\nonumber
 \eta_{\rm 144\,MHz}({\rm SFR}) & = & (0.83\pm 0.07) \left(\frac{\rm SFR}{\rm M_\sun\,yr^{-1}}\right )^{0.35\pm 0.07}\\
    \eta_{\rm 1365\,MHz}({\rm SFR}) & = & (1.38\pm 0.24) \left(\frac{\rm SFR}{\rm M_\sun\,yr^{-1}}\right )^{0.23\pm 0.04}.
\end{eqnarray}
With this we can expand the Condon-relation to correct for the CRE calorimetric efficiency
\begin{eqnarray}
\nonumber
\frac{\rm SFR_{\text{RC}}^\prime }{\rm M_\sun\,yr^{-1}} & = & \frac{1.22 \times 10^{-22}}{ \eta_{144\,\rm MHz}({\rm SFR})}\frac{L_{144\,\rm MHz}}{\rm W\,Hz^{-1}} \\
   \frac{\rm SFR_{\text{RC}}^\prime }{\rm M_\sun\,yr^{-1}}  & = & \frac{7.35 \times 10^{-22}}{\eta_{1365\,\rm MHz}({\rm SFR})}\frac{L_{1365\,\rm MHz}}{\rm W\,Hz^{-1}} .
\end{eqnarray}
to obtain our new calibration
\begin{eqnarray}
\nonumber
  \frac{\rm SFR_{\text{RC}}^\prime }{\rm M_\sun\,yr^{-1}} & = & (1.47\pm 0.12)   \frac{10^{-22} L_{144\,\rm MHz}/({\rm W\,Hz^{-1}})}{(L_{144\,\rm MHz}/L_{\rm N1})^{0.35\pm 0.07}}\\
    \frac{\rm SFR_{\text{RC}}^\prime }{\rm M_\sun\,yr^{-1}} & = & (0.54\pm 0.09)  \frac{10^{-21} L_{1365\,\rm MHz}/({\rm W\,Hz^{-1}})}{(L_{1365\,\rm MHz}/L_{\rm N2})^{0.23\pm 0.04}},
\end{eqnarray}
where $L_{\rm N1}=6.80\times 10^{21}\,\rm W\,Hz^{-1}$ and $L_{\rm N2}=1.84\times 10^{21}\,\rm W\,Hz^{-1}$ is the radio continuum luminosity of a galaxy with $\rm SFR=1\,\rm M_\sun\,yr^{-1}$ at 144 and 1365\,MHz, respectively. This calibration is valid for $0.1\lesssim {\rm SFR}/({\rm M_\sun/,yr^{-1}})\lesssim 10$ and for total masses within the star-forming disc of $10^9\lesssim M_{\rm tot}/{\rm M_\sun} \lesssim 10^{11}$. Our calibration at 1365\,MHz is in reasonable agreement with the prescription by \citet{bell_03a} who also took the non-linearity of the radio--SFR relation into account. 

One other point is that the slope of the relation can also change with respect
to the SFR tracer one uses \citep{bell_03a,boselli_15a}. This effect can be related to the large uncertainty in the data, to age dependencies on the SFR tracers, to the determination of SFR from the observed data once corrected for dust  attenuation. A question hopefully to be answered in the future is whether our simple model is still valid if one uses a different star formation tracer? Another question is the influence of the underlying assumption of the radio--SFR relation that the SFR has been constant over a few CRE radiative life-times. Otherwise, relatively fast variations in SFR would induce an excess or deficiency in the radio luminosity--SFR relation as tentatively shown \citet{ignesti_22a} for extreme jellyfish galaxies.

\subsection{Cosmic-ray electron transport }
We show in this work that the most likely transport mechanism for CREs in our galaxy sample is energy-independent diffusion. This is in agreement with our previous results for CRE diffusion in the disc of NGC~5194 where we analysed five frequencies at 54--8350\,MHz \citep{heesen_23b}. In \citet{doerner_23a}, we showed that 3D simulations of energy-independent diffusion with advection in a galactic wind can explain the radio continuum data of this galaxy. Also, in edge-on galaxies we do find similar results where the diffusion coefficient is energy-independent in those galaxies that do not have a galactic wind \citep{heesen_22a,stein_23a}.

Hence, in summary we find that CRE transport is energy independent in the energy range of a few GeV (<10\,GeV). Theoretical explanations for energy-independent diffusion include field line random walk, where the cosmic rays simply follow the magnetic field lines that are randomised and bent by turbulence \citep{minnie_09a}; alternatively, cosmic rays are scattered in turbulence that they generate themselves in the so-called self-confinement picture \citep{zweibel_13a}; lastly, cosmic ray may be scattered in extrinsic turbulence that is generated in the interstellar medium such as by supernovae and stellar winds. See \citet{heesen_23b} for more information and references.

Any theory has to be able to explain why there appear to be two types of galaxies namely the ones with diffusive haloes and the ones with advective haloes. Galaxies that have sufficiently high SFR surface densities can possess galactic winds that may be even cosmic ray-driven \citep{breitschwerdt_91a,everett_08a}. On the other hand, galaxies that are rather quiescent do not have galaxy-wide winds so that the cosmic-ray transport appears of diffusive fashion. The self-confinement picture would be able to do this \citep{zweibel_13a}, but also more diffusive transport modes are able to drive galactic winds \citep{wiener_17a}.

While our results appear to be at first glance in contradiction to some Earth-based measurements, we need to keep in mind the different energy range. The observed change of primary-to-secondary cosmic rays that suggest an energy-dependent diffusion coefficient \citep{becker_tjus_20a}, apply to energies of $>$10\,GeV. In contrast, our data probe lower energy CRE with energies of less than 10\,GeV, whereas the solar wind prevents this kind of measurement near the Earth. In this energy range there appears to be a flattening of the cosmic-ray spectrum as suggested by PAMELA and more importantly by {\it Voyager} data, which could point to a similar change of transport mode \citep{blasi_12a}.

\section{Conclusions}
\label{sec:conclusions}

Radio continuum emission can be used as an extinction-free star formation tracer. This is motivated by the existence of a radio--SFR relation both globally, between galaxies, and spatially resolved, within galaxies. The link between radio continuum emissions and star formation are CREs and their interactions with the ambient magnetic field. We want to understand the mechanism of CRE transport and spectral ageing using their radio spectral indices in order to calibrate the local radio--SFR relation. Such a calibration would allow us to measure SFRs in galaxies at higher redshifts with future surveys such as with the Square Kilometre Array. We investigated a sample of 15 nearby face--on galaxies with data from WSRT, MeerKAT, SINGS and LOFAR at two different radio frequencies of 144 and 1365\,MHz. Low frequency data are almost uncontaminated by thermal (free--free) emission. We follow the approach by \citet{heesen_14a} where  the radio--SFR relation is measured at a spatial resolution of (1.2\,kpc)$^2$ and the radio continuum emission is compared with SFR surface densities using a hybrid prescription of far-ultraviolet and mid-infrared data \citep{leroy_08a,leroy_12a}.

In agreement with earlier studies \citep[e.g.][]{berkhuijsen_13a,vollmer_20a} we find the SFR maps are smeared-out versions of the radio maps due to CRE transport resulting in sublinear radio--SFR relations. We linearise the radio--SFR relation by convolving the SFR maps with a Gaussian kernel, where the kernel size defines the CRE transport length. We find that the spatially resolved radio--SFR relations have sublinear slopes that move closer to linearity with increasing frequency (Fig.\,\ref{fig:radio_sfr}a,b). The trend has also been found in other studies \citep[e.g.][]{heesen_19a}  and supports the idea of CRE transport and spectral ageing. This is corroborated by the radio--SFR relation being nearly linear if we restrict our study to areas with young CRE (Fig.\,\ref{fig:radio_sfr}c,d). We now investigated the deviation of the local radio--SFR relation as a function of radio spectral index (Fig.\,\ref{fig:radio_sfr_alpha}). The ratio of radio-to-hybrid SFR surface density is then defined as the CRE calorimetric efficiency (Eq.\,\ref{eq:calorimetric_efficiency}) which can be parametrised as function of the radio spectral index (Eq.\,\ref{eq:calorimetric_sfr}). Correcting the radio--SFR relation for the calorimetric efficiency, Figure\,\ref{fig:radio_sfr_calor} presents our main result. The corrected radio--SFR relation is now almost in agreement with a linear relation.

The smoothing experiment shows that lower energy CREs traced by lower frequencies have longer transport lengths that can not exceed about half of the radius of the star-forming disc (Fig.\,\ref{fig:len_radius}). The CRE diffusion length has weakly significant correlation with SFR surface density as shown in Fig.\,\ref{fig:len_sfrd}. This was already found by \citet{murphy_08a} and is interpreted by higher magnetic field strengths in areas of higher SFR \citep{heesen_23a,pfrommer_22a}. This is corroborated by a weak correlation between CRE transport length and lifetime (Fig.\,\ref{fig:len_tau}). However, there is no significant correlation between CRE diffusion length and total magnetic field strength (Fig.\ref{fig:len_bfield}), so that the CRE is not the only influence that determines the transport length. The magnetic field structure is likely of importance too. The relation between CRE transport length and lifetime shown in Fig.\,\ref{fig:len_tau} agrees best with energy-independent diffusion. This is also the case when we consider the ratio of CRE transport length at 144 to that at 1365\,MHz (Tab.\,\ref{tab:diff}). This is in agreement with earlier studies \citep{heesen_23b,doerner_23a} and applies to low-energy GeV CREs. We note that our observations of CRE transport apply only to a globalised measurement on kpc-scales, where there will be some confusion with the halo emission \citep{stein_23a}.

Arguably the most exciting aspect of our study is the application of the corrected radio--SFR relation to integrated measurements. If we parametrise the calorimetric efficiency with the radio spectral index and apply this as a correcting factor to the integrated radio--SFR relation of the same LoTSS-DR2 galaxies \citep{heesen_22a}, we find the same linear radio--SFR relation as for the spatially resolved measurement (Fig.\,\ref{fig:integrated_radio_sfr}). As the normalisation of the integrated radio--SFR relation (Eq.\,\ref{eq:integrated_radio_sfr}) agrees with the spatially resolved one (Eq.\,\ref{eq:condon}), we have now identified a potentially universal radio--SFR relation that applies to both global and local measurements. In particular, we are now also able to correct global measurements for the CRE calorimetric efficiency with the help of the radio spectral index. This means that we may be able to reduce in the future the systematic uncertainties of the radio--SFR relation that have thus far limited the applicability of deep radio continuum surveys to measure cosmic SFRs as will become possible with future surveys with LOFAR, MeerKat, and the SKA.



\begin{acknowledgement}
We thank the anonymous referee for a constructive report that helped to improve the clarity of the paper. LOFAR \citep{vanHaarlem_13a} is the Low Frequency Array designed and constructed by ASTRON. It has observing, data processing, and data storage facilities in several countries, that are owned by various parties (each with their own funding sources), and that are collectively operated by the ILT foundation under a joint scientific policy. The ILT resources have benefitted from the following recent major funding sources: CNRS-INSU, Observatoire de Paris and Universit\'e d'Orl\'eans, France; BMBF, MIWF-NRW, MPG, Germany; Science Foundation Ireland (SFI), Department of Business, Enterprise and Innovation (DBEI), Ireland; NWO, The Netherlands; The Science and Technology Facilities Council, UK; Ministry of Science and Higher Education, Poland.

M.B. acknowledges funding by the Deutsche Forschungsgemeinschaft (DFG, German Research Foundation) under Germany’s Excellence Strategy – EXC 2121 ‘Quantum Universe’ – 390833306. M.S. acknowledges funding from the German Science Foundation DFG, within the Collaborative Research Center SFB1491. This research has made use of the NASA/IPAC Extragalactic Database (NED),
which is operated by the Jet Propulsion Laboratory, California Institute of Technology,
under contract with the National Aeronautics and Space Administration.
This work made use of the {\sc SciPy} project \href{https://scipy.org}{https://scipy.org}. 

\end{acknowledgement}

%
%

\bibliographystyle{aa}
\bibliography{review} 



    \label{fig:stellar}




\end{document}